\documentclass[twocolumn,aps,prd,superscriptaddress,nofootinbib,preprintnumbers,showpacs]{revtex4-1}

\usepackage{bm,amsmath,amsfonts,amssymb,graphicx,diagbox}

\usepackage{breakurl}
\usepackage{array}
\usepackage{slashed}
\usepackage{xspace}
\usepackage{booktabs}
\usepackage{hhline}
\usepackage{multirow}
\usepackage{header3}
\usepackage{placeins}
\usepackage{physics}

\allowdisplaybreaks
\newcommand{\RRho}{R(\rho)}
\newcommand{\ROmega}{R(\omega)}
\newcommand{\RV}{R(V)}
\newcommand{\redRRho}{\widetilde{R}(\rho)}
\newcommand{\redROmega}{\widetilde{R}(\omega)}
\newcommand{\redRV}{\widetilde{R}(V)}

\newcommand{\BtoPiLNu}{B \to \pi l \bar\nu}
\newcommand{\BtoPiTauNu}{B \to \pi \tau \bar\nu}
\newcommand{\BtoRhoLNu}{B \to  \rho l \bar\nu}
\newcommand{\BtoRhoTauNu}{B \to  \rho \tau \bar\nu}
\newcommand{\BtoOmegaLNu}{B \to  \omega l \bar\nu}
\newcommand{\BtoOmegaTauNu}{B \to \omega \tau \bar\nu}
\newcommand{\BtoVLNu}{B \to  V l \bar\nu}
\newcommand{\BtoVTauNu}{B \to  V \tau \bar\nu}

\newcommand\mat[1]{\mathbf{#1}}
\renewcommand\vec[1]{\mathbf{#1}}

\newcommand{\nn}{\nonumber}
\newcommand{\GeV}{\text{GeV}}

\newcommand{\babar}{\mbox{\ensuremath{{\displaystyle B}\!{\scriptstyle A}{\displaystyle B}\!{\scriptstyle AR}}}\xspace}

\newcommand{\mn}{{\mu\nu}}
\newcommand{\g}{\gamma}

\newcommand{\bbar}{\bar{b}}

\newcommand{\ubar}{\bar{u}}

\newcommand{\Bbar}{\,\overline{\!B}{}}
\newcommand{\Dbar}{\,\overline{\!D}{}}
\newcommand{\lqcd}{\ensuremath{\Lambda_{\rm QCD}}\xspace}

\newcommand{\Vub}{V_{\text{ub}}}
\renewcommand{\Vcb}{V_{\text{cb}}}
\newcommand{\dotpr}[2]{\,#1\!\cdot #2\, }

\newcommand{\ampBb}[2]{\big\langle #1 \big|\, #2\, \big| \Bbar \big\rangle}
\renewcommand{\ampBb}[2]{\langle #1 |\, #2\, | \Bbar \rangle}

\def\spnt{1}
\if\spnt1
	\newcommand{\up}{+}
	\newcommand{\dn}{-}
\else
	\newcommand{\up}{2}
	\newcommand{\dn}{1}
\fi

\def\SR{S\!R}
\def\SL{S\!L}
\def\VR{V\!R}
\def\VL{V\!L}
\newcommand{\alSL}{\alpha_L^S}
\newcommand{\alSR}{\alpha_R^S}
\newcommand{\alVL}{\alpha_L^V}
\newcommand{\alVR}{\alpha_R^V}
\newcommand{\alTL}{\alpha_L^T}
\newcommand{\alTR}{\alpha_R^T}
\newcommand{\beSL}{\beta_L^S}
\newcommand{\beSR}{\beta_R^S}
\newcommand{\beVL}{\beta_L^V}
\newcommand{\beVR}{\beta_R^V}
\newcommand{\beTL}{\beta_L^T}
\newcommand{\beTR}{\beta_R^T}

\newcommand{\rU}{r_{V}}
\newcommand{\rl}{r_l}
\newcommand{\mSqq}{\hat q^2}
\newcommand{\mPw}{|\bar{p}_V\!|}

\newcommand{\thtau}{\theta_{l}}
\newcommand{\phtau}{\phi_{l}}

\newcommand{\AP}{A_{P}}
\newcommand{\FV}{V}
\newcommand{\FA}[1]{A_{#1}}
\newcommand{\FT}[1]{T_{#1}}

\makeatletter
\g@addto@macro\bfseries{\boldmath}
\makeatother

\definecolor{nicered}{rgb}{0.7,0.1,0.1}
\definecolor{nicegreen}{rgb}{0.1,0.5,0.1}
\definecolor{niceblue}{rgb}{0.1,0.1,0.5}

\usepackage[bookmarks=false]{hyperref}
\hypersetup{colorlinks,citecolor= nicegreen,linkcolor= niceblue,urlcolor= niceblue}

\allowdisplaybreaks

\begin{document}

\author{Florian~U.~Bernlochner}
\author{Markus~T.~Prim}
\affiliation{Physikalisches Institut der Rheinischen Friedrich-Wilhelms-Universit\"at Bonn,\\53115 Bonn, Germany\\[4pt]}

\author{Dean~J.~Robinson}
\affiliation{Ernest Orlando Lawrence Berkeley National Laboratory, University of California,\\ Berkeley, CA 94720, USA\\[4pt]}

\title{$B \to \rho l \bar \nu$ and $\omega l \bar \nu$ in and beyond the Standard Model: Improved predictions and $|V_{ub}|$}

\begin{abstract}
We revisit the experimental and theoretical status of $B \to \rho l \bar{\nu}$ and $B \to \omega l \bar{\nu}$ decays.
We perform a combined fit of averaged spectra from Belle and Babar measurements with prior light cone sum rule calculations,
in order to obtain more precise predictions over the full $q^2$ range.
The extracted values of $|V_{ub}|$ from these combined fits exhibit smaller uncertainty compared to previous extractions from  $B \to \rho l \bar{\nu}$ and $B \to \omega l \bar{\nu}$ decays and the central values are found to be smaller than values extracted from $B \to \pi l \nu$ or inclusive measurements.
We use our fit results to obtain more precise predictions in and beyond the Standard Model for the lepton universality ratios $R(\rho)$ and $R(\omega)$, 
as well as several angular observables that are sensitive to the full $q^2$ distribution, 
such as the longitudinal polarization of the vector meson, the $\tau$ polarization, and its forward-backward asymmetry. 
\end{abstract}

\maketitle

\section{Introduction}

Semileptonic decays offer a clean laboratory to search for physical phenomena beyond those predicted by the Standard Model (SM) of particle physics. 
Intriguingly, semitauonic transitions involving charmed final states---i.e., $b \to c \tau \bar\nu$ decays---show 
a persistent Lepton Flavor Universality Violation (LFUV) anomaly at the $3\,\sigma$ level~\cite{Amhis:2019ckw}, 
or higher, when various decay modes are combined; see Ref.~\cite{Bernlochner:2021vlv} for a recent review. 
As pointed out by e.g. Refs.~\cite{Bernlochner:2015mya,Becirevic:2020rzi,Leljak:2021vte}, semitauonic transitions involving 
charmless hadronic final states offer an intriguing independent probe of LFUV anomalies. 
In particular, exploring $b \to u \tau \bar\nu$ decays can be a sensitive probe of the flavor structure of New Physics (NP) mediators (if any) responsible for the $b \to c\tau\bar\nu$ LFUV anomalies.
For instance, if the same NP were present in $b \to u$ semitauonic decays as in $b \to c$ semitauonic decays, 
one would naively expect LFUV deviations from the SM in the former to be enhanced by $|\Vcb|^2/|\Vub|^2 \sim 10^2$,
compared to the $\sim 10$--$20$\% LFUV excess rates seen in $b \to c\tau\bar\nu$.

In 2015, Belle published the first search for $B \to \pi \tau \bar \nu$ using single-prong hadronic and leptonic $\tau$ decays~\cite{Hamer:2015jsa}. 
Their measured upper limit of the branching fraction can be translated into a CL for the ratio of semitauonic and light-lepton modes~\cite{Bernlochner:2015mya}, 
i.e. the lepton universality ratio
\begin{equation}
 	R(\pi) = \frac{\Gamma(B \to \pi \tau \bar \nu)}{\Gamma(B \to \pi \ell \bar \nu)} = 1.05 \pm 0.51 \, ,
\end{equation}
with $\ell = e$ or $\mu$.
The measured value is compatible with the SM predictions $R(\pi)_{\text{SM}} = 0.641 \pm 0.016$~\cite{Bernlochner:2015mya} or $0.688 \pm 0.014$~\cite{Leljak:2021vte}.  
It is expected that Belle~II will discover this decay, and then push its measured precision to the $5$--$6\%$ level with the anticipated full data set~\cite{Bernlochner:2021vlv}. 

In this paper, we explore  $\BtoRhoTauNu$ and $\BtoOmegaTauNu$ transitions (collectively denoted $\BtoVTauNu$). 
Measurements of $\BtoVTauNu$ decays feature several advantages over $\BtoPiTauNu$ in terms of their potential sensitivity to NP effects, including
an increased branching fraction with respect to the pion final state,
and a larger set of angular observables from the subsequent $\rho \to \pi\pi$ and $\omega \to \pi\pi\pi$ decays,
that may probe NP effects arising in the polarization of the $\rho$ and $\omega$.
The Belle~II experiment has started recording its first collision data:
Large and clean data sets of these final states will soon be available to probe the full differential information in these decays.
In addition, LHCb has established itself as a source of precise measurements of semileptonic processes. 
With its sizeable data sets, it is conceivable that its first measurements of $b \to u \tau \bar\nu$ transitions will appear in the near future, 
with $B^+ \to \rho^0 \, \tau^+ \, \bar \nu$ and $\Lambda_b \to p \, \tau \, \bar \nu$ being likely candidates. 

To produce reliable $\BtoVTauNu$ predictions for both the SM and NP, in this paper we reanalyze the available experimental measurements of the differential decay rates
in $q^2$ for $B \to \rho \ell \bar \nu$ and $B \to \omega \ell \bar \nu$, published by \babar\ and Belle in Refs~\cite{Sibidanov:2013rkk,delAmoSanchez:2010af,Lees:2012mq}. 
Newly-averaged spectra are obtained, following the prescription utilized by HFLAV for $B \to \pi \ell \bar \nu$~\cite{Amhis:2019ckw}. 
These spectra are then fitted simultaneously with light-cone sum rule (LCSR) predictions from Ref.~\cite{Straub:2015ica}.
This generates improved fit results for its particular parameterization of the SM and NP form factors, which we refer to as the `BSZ' parametrization hereafter.
Our combination of LCSR and experimental information extends the applicability of fits to the BSZ parametrization to the full $q^2$ range.
This not only allows for more reliable predictions of observables that are sensitive to integrations over the full phase-space, but also for the use of the entire $q^2$ range to determine $|\Vub|$,  instead of only the low $q^2$ region . We thus redetermine the CKM matrix element $|\Vub|$, and compare it to the values from Refs.~\cite{Amhis:2019ckw,Leljak:2021vte}, 
which were determined from exclusive $B \to \pi \ell \bar \nu$, and to the recent result from inclusive $b \to u \ell \bar \nu$ decays~\cite{Cao:2021xqf}. 

Using our combined fit results, we provide improved SM predictions for the lepton universality ratio $R(V)$ and several angular observables, such as the longitudinal polarization of the vector meson,
the $\tau$ polarization, and its forward-backward asymmetry. 
(In doing so, we also provide the explicit construction of the $\omega$ longitudinal polarization in terms of the $B \to (\omega \to \pi\pi\pi) \tau \bar\nu$ 5-body differential rate.)
We further briefly explore the potential to search for NP effects in $\BtoRhoTauNu$ and $\BtoOmegaTauNu$ transitions.
Just as for the charmed final states, forward-folded model-independent approaches that exploit the full differential information to fit directly to the NP Wilson coefficients, 
may be required to avoid biases in NP interpretations of unfolded observables such as $R(V)$~\cite{Bernlochner:2020tfi}, should an anomaly be seen.
Such model-independent Wilson Coefficient fits 
can naturally be applied to charmless final states, 
in order to constrain the NP model space in a model-independent manner. 

This paper is organized as follows.
Section~\ref{sec:theory} introduces the relevant theoretical foundations and conventions used to describe $b \to u l\bar\nu$ decays ($l = e$, $\mu$ or $\tau$).
This includes the explicit construction of the $B \to V$ form factors as well as their parametrization with respect to LCSR results~\cite{Straub:2015ica},
along with expressions for the $\BtoVLNu$ differential decay rates. 
Expressions for the $B \to (\rho \to \pi\pi) l \bar\nu$ and $B \to (\omega \to \pi\pi\pi) l \bar\nu$ NP amplitudes and differential rates, are provided in an Appendix.  
In Sec.~\ref{sec:average}, we derive an averaged $q^2$ spectrum combining experimental results from both \babar\ and Belle,
and Sec.~\ref{sec:theo_fit} proceeds to discuss our combined experimental plus LCSR fit to determine the form factors. 
In Sec.~\ref{sec:preds}, we present improved SM predictions for various observables, followed by a brief discussion of NP effects.

\section{$B \to V l \bar\nu$ in the SM and beyond}
\label{sec:theory}

The effective Standard Model (SM) Lagrangian describing semileptonic $b \to u$ transitions arises from the four-Fermi interaction
\begin{equation}
	\mathcal{L}^\mathrm{SM}_\mathrm{eff} = \frac{-4G_F}{\sqrt{2}} \Vub (\bar{u} \g_\mu P_L b) (\bar{l}  \g^\mu P_L \nu) + \mathrm{h.c.},
\end{equation}
in which $\Vub$ is the CKM matrix element, the chiral projectors $P_{R,L}= (1\pm \g_5)/2$, and the Fermi constant $G_F^{-1}= 8 m_{W}^2 / (\sqrt{2}g_2^2)$, 
with $m_{W}$ the $W$ mass and $g_2$ the electroweak $SU(2)_{L}$ coupling constant.
Throughout this paper, we denote the light leptons by $\ell = e$ or $\mu$, while $l = e$, $\mu$ or $\tau$.
The parton level $b \to u l \bar \nu$ decay amplitude in the SM is given by
\begin{equation}
	\mathcal{A} = \quad \quad \begin{gathered}
	\includegraphics[width=3.5cm]{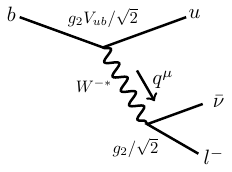}
	\end{gathered}
\end{equation}
The quarks are dressed into different hadrons, that participate in various exclusive decay modes. 
In particular, we focus on $\BtoVLNu$  decays, with vector meson $V = \rho$ or $\omega$.

Anticipating a discussion of New Physics later on, a generalized version of the Lagrangian including arbitrary NP contributions to $b \to u \tau\bar\nu$ can be written as
\begin{equation}
	\label{eqn:LeffNP}
	\mathcal{L_\mathrm{eff}} = \mathcal{L}^\mathrm{SM}_\mathrm{eff} - \frac{c_{XY}}{\Lambda_{\text{eff}}^2} (\bar{u} \Gamma_{X} b) (\bar\tau \Gamma_Y \nu) + \mathrm{h.c.}.
\end{equation}
Here $\Gamma_{X(Y)}$ is any Dirac matrix and $c_{XY}$ is the corresponding NP Wilson coefficient defined at scale $\mu \sim m_b$.
(We assume that NP only affects the $b \to u \tau \bar \nu$ decays, and not the light-lepton modes.)
In Eq.~\eqref{eqn:LeffNP} we have normalized the NP Wilson coefficients with respect to the SM current, 
such that the effective scale $\Lambda_{\text{eff}} = [4G_F/\sqrt{2} \Vub]^{-1/2} \simeq 2.7$\,TeV.
Our choice of basis for $\Gamma_{X}$ is the set of chiral scalar, vector and tensor currents. 
That is, $\Gamma_{x} = P_{R,L}$, $\g^\mu P_{R,L}$, and $\sigma^{\mu\nu} P_{R,L}$, respectively.
Assuming only SM left-handed neutrinos, the lepton current is always left-handed, and the tensor quark current may only be left-handed.
We write the five remaining Wilson Coefficients as $c_{XY} = c_{\SR}$, $c_{\SL}$, $c_{\VR}$, $c_{\VL}$ and $c_T$.\footnote{
In contexts that one considers also right-handed neutrinos, an alternative (slightly abused) notation is
$c_{XY} = S_{qXlY}$, $V_{qXlY}$, $T_{qXlY}$, where the $S$, $V$, $T$ denotes the Lorentz structure of the quark and lepton currents, 
and $X$, $Y$ = $L$, $R$ instead denotes the chirality of the $b$ quark or charged lepton, respectively.
This is the notation used in e.g. the \texttt{Hammer} library~\cite{Bernlochner:2020tfi,bernlochner_florian_urs_2020_3993770}.
The explicit correspondence between the two choices is:
$c_{\SR} = S_{qRlL}$, $c_{\SL} = S_{qLlL}$, $c_{\VR} = V_{qRlL}$, $c_{\VL} = V_{qLlL}$, and $c_T = T_{qLlL}$.
}

The hadronic matrix elements arising in exclusive $B \to V$ transitions can be generically written as
\begin{equation}
	\mel{V(p_V)}{ \bar{u} \, \Gamma\, b }{\Bbar(p_{B})} = c_V \sum_i \mathcal{T}^\Gamma_i F^\Gamma_i(q^2),
	\label{eq:tensor-decomposition}
\end{equation}
where $q=p_{B}-p_V$. The Clebsch-Gordan coefficient takes the value $c_V = 1/\sqrt{2}$ for the neutral unflavored meson final states $V = \rho^0$ and $\omega^0$, while $c_V = 1$ for $\rho^\pm$. 
For each current $\ubar \Gamma b$, the $\mathcal{T}^\Gamma_i$ denote a basis of the allowed amplitudes---tensors of the involved 4-momenta and polarizations---while 
$F^\Gamma_i$ are their corresponding form factors. 
In $B \to V$ transitions there are a total of eight possible independent amplitudes, and hence eight form factors.
As in Ref.~\cite{Straub:2015ica} we choose the basis $\{A_P, V, A_0, A_1, A_{12}, T_1, T_2, T_{23}\}$
defined explicitly via\footnote{It is perhaps unfortunate that the notation for the vector form factor $V = V(q^2)$ is identical to the notation for the vector meson, $V$. 
Which is meant will always be clear from context.}
\begin{subequations}
\label{eqn:FFcurr}
	\begin{align}
		& \ampBb{V}{\ubar \g^5 b}  = c_V A_P \dotpr{\varepsilon^*}{q} \,,\\
		&\ampBb{V}{\ubar \g^\mu b}  =  \frac{i c_V V  \epsilon^{\mn \rho \sigma}\, \varepsilon^*_\nu\, (p_B + p_V)_\rho\, q_\sigma}{m_B + m_V}\,, \label{eqn:Vcurr} \\
		& \ampBb{V}{\ubar \g^\mu \g^5 b} =  c_V\bigg\{ A_1(m_B + m_V) \varepsilon^{*\mu} \nn \\
		& \qquad - A_2\frac{(p_B + p_V)^\mu \dotpr{\varepsilon^*}{q}}{m_B + m_V} + \frac{\dotpr{\varepsilon^*}{q} q^\mu}{q^2}\Big[ A_2(m_B - m_V) \nn \\
		& \qquad - A_1(m_B + m_V) + 2 m_V A_0 \Big] \bigg\}\,, \label{eqn:Acurr}\\
		& \ampBb{V}{\ubar \sigma^\mn b}  =  -c_V \epsilon^{\mn \rho \sigma}\bigg\{ T_1 \varepsilon^*_\rho (p_B + p_V)_\sigma \label{eqn:Tcurr} \\
		& \qquad - (T_2 + T_1)\frac{m_B^2 - m_V^2}{q^2}\varepsilon^*_\rho q_\sigma \nn \\
		& \qquad + (p_B + p_V)_\rho q_\sigma \frac{\dotpr{\varepsilon^*}{q}}{q^2} \bigg[(T_1 + T_2)  + T_3 \frac{q^2}{m_B^2 - m_V^2}\bigg]\bigg\} \,,\nn
	\end{align}
\end{subequations} 
with the additional redefinitions with respect to $A_{12}$ and $T_{23}$,
	\begin{subequations}
		\begin{align}
		4|p_V|^2 m_B^2 A_2 & = A_1(m_B^2 - m_V^2 -q^2)(m_B + m_V)^2  \nn \\
		& \qquad - 16 A_{12} m_B m_V^2(m_B + m_V)\,,\\
		4|p_V|^2 m_B^2T_3 & = T_2(m_B^2 + 3m_V^2 -q^2)(m_B^2 - m_V^2) \nn \\
		& \qquad - 8 T_{23} m_B m_V^2(m_B - m_V)\,.
		\end{align}
	\end{subequations}
Here $m_B$ ($m_V$) is the mass of the $B$ (vector) meson, and $|p_V|$ denotes the vector meson 3-momentum in the $B$ rest frame, 
\begin{align}
	|p_V| & = m_V\sqrt{w^2-1}\,, \nn \\
	& = \sqrt{\lambda(q^2)}/(2 m_B)\,, \qquad w = \frac{m_B^2 + m_V^2 - q^2}{2m_B m_V} 
\end{align}
in which the K\"allen function $\lambda(q^2) = [(m_{B} + m_{V})^2 - q^2][(m_{B} - m_{V})^2 - q^2]$. Note $\ampBb{V}{\ubar b} = 0$ by angular momentum and parity conservation.
The identity $\sigma^{\mu\nu} \g^5 \equiv -(i/2)\, \varepsilon^{\mu\nu \rho \sigma} \sigma_{\rho \sigma}$, 
corresponding to $\text{Tr}[\g^\mu\g^\nu\g^\sigma\g^\rho\g^5] = +4i \varepsilon^{\mu\nu\rho\sigma}$,
allows one to write down the matrix element for the axial-tensor current $\ampBb{V}{\ubar \sigma^\mn \g^5 b}$ from the tensor~\eqref{eqn:Tcurr}.
This is the standard Lorentz sign convention in the $B \to D^{*}$ literature. 
One may instead choose the sign conventions such that $\sigma^{\mu\nu} \g^5 \equiv +(i/2)\, \varepsilon^{\mu\nu \rho \sigma} \sigma_{\rho \sigma}$,
corresponding to the more common $\text{Tr}[\g^\mu\g^\nu\g^\sigma\g^\rho\g^5] = -4i \varepsilon^{\mu\nu\rho\sigma}$.
In this case the sign of the vector and tensor currents in Eqs.~\eqref{eqn:Vcurr} and~\eqref{eqn:Tcurr} also changes.

The construction of the form factor basis in Eqs.~\eqref{eqn:FFcurr} assumes the vector meson $V$ may be treated as an on-shell state. 
While a good assumption for the narrow $\omega^0$, this is a poorer assumption for the relatively broad $\rho$, cf. Ref.~\cite{Kang:2013jaa}. 
For instance, once subsequent $\rho \to \pi\pi$ decays are considered, longitudinal modes may generate important contributions naively $\sim (1 - p_V^2/m_V^2) \sim \Gamma_V/m_V$.
However, in a sufficiently narrow range of $p_V^2$ near the $\rho$ pole, such effects are always subleading, albeit at the expense of a smaller branching ratio,
and we shall therefore not consider these effects further here. 
Note that constraining the broad resonance to this narrow range also fixes the endpoint of the $q^2$ range to (be close to) the usual $(m_B - m_V)^2$, with $m_V$ the pole mass of the vector meson.
We do not discuss the impact of non-resonant $B \to \pi \pi \ell \bar \nu_\ell$ production on the predicted rates:
The impact of such contributions in light leptons was recently measured for the first time in Ref.~\cite{Beleno:2020gzt}, 
which reported the full sum of resonant and non-resonant semileptonic di-pion final states, 
i.e. the sum of $\rho$, higher resonant states, as well as non-resonant $\pi\pi$ contributions. 

In Appendix~\ref{sec:amplrates} we provide the explicit forms of the $\BtoVLNu$ helicity amplitudes for all SM and NP couplings, 
as well as the amplitudes and full differential rates once subsequent $\rho \to \pi\pi$ or $\omega \to \pi\pi\pi$ decays are included.
For the purposes of our fit below, it is enough to present here just the SM amplitudes and differential rate for $\BtoVLNu$.
NP effects are discussed further in Sec.~\ref{sec:preds}.
In the standard helicity basis, the $\BtoVLNu$ helicity amplitudes take the form (up to an overall unphysical phase; see App.~\ref{sec:amplrates})
\begin{subequations}
\begin{align}
	H_\pm(q^2) &=  \frac{2m_B |p_V| V(q^2)}{m_{B} + m_V} \pm (m_{B} + m_V) A_1(q^2) \, ,\\
	H_0(q^2) &= 8m_{B}m_V A_{12}(q^2)/\sqrt{q^2} \, ,\\
	H_s(q^2) &= 2m_B|p_V|A_0(q^2)/\sqrt{q^2} \, .
\end{align}
\end{subequations}
The SM differential rate is then given by
\begin{align}
	 \frac{\mathrm{d}\Gamma}{\mathrm{dq^2}}  & =  \frac{G_F^2\big| \Vub \big|^2 c_V^2}{96 \pi^3} |p_V| \frac{q^2}{m_B^2} \bigg( 1- \frac{m_{l}^2}{q^2} \bigg)^2 \nn \\
	 & \quad \times \bigg\{ \bigg[ 1+\frac{m_l^2}{2q^2} \bigg]\Big(H_+^2(q^2) + H_-^2(q^2) + H_0^2(q^2) \Big) \nn \\
	 & \qquad \qquad + \frac{3 m_l^2}{2q^2}  H_s^2(q^2) \bigg\}\,.\label{eq:ulnu-theory:BToVlnuFull}
\end{align}
In line with the approach of the $B \to \pi \ell \bar \nu$ analysis of Ref.~\cite{Amhis:2019ckw}, 
the electroweak correction~\cite{Sirlin:1981ie} for semileptonic decays, $\eta_\mathrm{EW} = 1+ (\alpha/\pi) \log (m_Z/m_B) \approx 1.0066$, is not included in the rate. 
This correction can always be applied \emph{post facto} using the transformation $|\Vub| \to |\Vub|\eta_\mathrm{EW}$. 
Additional long-distance QED corrections may further affect the determined value of $|\Vub|$.
These corrections have been estimated to be small using a scalar QED approximation with some model assumptions~\cite{Tostado:2015tna}. 
In the massless lepton limit, the scalar helicity amplitude $H_s$, and hence $A_0$, does not contribute, reducing the SM form factors to three. 
The `zero mass approximation'---neglecting the electron or muon mass---is used in Secs.~\ref{sec:legacy-spectrum} and~\ref{sec:theo_fit} below to obtain fits for $V$, $A_1$, and $A_{12}$.

The form factors themselves are hadronic functions that cannot be determined with perturbative methods, since they incorporate non-perturbative QCD effects. 
However, one may exploit dispersion relations plus analyticity and unitarity bounds to parametrize them in a model-independent manner. 
Similarly to the BGL parametrization for $B \to D^{(*)}$~\cite{Boyd:1995sq,Boyd:1997kz}, 
the Bourrely-Caprini-Lellouch (BCL) parametrization~\cite{Bourrely:2008za} 
exploits a dispersive approach to express the (originally $B \to \pi$) form factors as a power expansion
with respect to the conformal parameter
\begin{equation}
	z(q^2, t_0) = \frac{\sqrt{t_+ - q^2} - \sqrt{t_+ - t_0^{\phantom{2}}}}{\sqrt{t_+ - q^2} + \sqrt{t_+ - t_0^{\phantom{2}}}} \,.
\end{equation}	
Here the pair production threshold $t_+ = (m_{B} + m_V)^2$ and the $z$-origin is determined by the optimized choice 
$t_0 = (m_{B} + m_V)(\sqrt{m_{B}} - \sqrt{m_V})^2$, that minimizes the range of $|z|$ to be $\le 0.10$.
(BCL further applies a constraint on the gradient of the $B \to \pi$ vector form factor at $z=-1$.)
Naive regularization of the $1/q^2$ terms in Eqs.~\eqref{eqn:Acurr} and~\eqref{eqn:Tcurr} imply the kinematic relations at $q^2 = 0$
\begin{equation}
	\label{eqn:q20rel}
	A_0(0) = \frac{8 m_B m_V A_{12}(0)}{m_B^2 - m_V^2}\,,\quad
	T_1(0) = T_2(0)\,.
\end{equation}
The Bharucha-Straub-Zwicky (BSZ) parametrization~\cite{Straub:2015ica} modifies the BCL parametrization, 
by reorganizing the power expansion in $z$ as a `simplified series expansion' about $q^2=0$,
in order to straighforwardly impose these relations at zeroth order. 
That is, the form factors are expanded as
\begin{equation}
	F_i(q^2) = P_i(q^2)\sum_k \alpha_k^i \Big(z(q^2) - z(0)\Big)^k\,.
\end{equation}
Just as for BCL, for each current a single subthreshold resonance is assumed at $q^2 = m_{R}^2$, 
such that the (inverse) Blaschke factor $P_i(q^2) = (1 - q^2/m_{R}^2)^{-1}$.
As allowed by angular momentum and parity, these resonances are explicitly for each of the form factors
\begin{align*}
	A_P\,,~A_0 &: \quad R = B, & m_B \simeq 5.279\,\GeV\,, \\
	V\,,~ T_1 &: \quad R = B^*, & m_{B^*} \simeq 5.325\,\GeV\,, \\
	A_{1,12}\,,~T_{2,23} &: \quad R = B_1, & m_{B_1} \simeq 5.724\,\GeV\,,
\end{align*}
coupling to $J^P = 0^-$, $1^-$ and $1^+$ partial waves, respectively.
Finally, the quark equations of motion may be used to relate the pseudoscalar form factor $A_P$ to $A_0$, via
\begin{equation}
	A_P = - \frac{2m_V}{m_{b} + m_{u}} A_0 \,.
\end{equation}
Here $m_{b,u}$ are formally scheme-dependent quantities. 
The BSZ parametrization~\cite{Straub:2015ica} uses the pole mass scheme, with explicitly $m_b \simeq 4.8$\,GeV,
and the much lighter $u$ quark mass is neglected. We use the same scheme.

Because of the unstable nature of the $\rho$ and $\omega$ mesons, lattice QCD (LQCD) predictions are challenging, 
and so far have not yet provided predictions with controlled systematic uncertainties that may be used in fits with data~\cite{Aoki:2019cca}.
One may instead exploit light-cone sum rule (LCSR)~\cite{BALITSKY1989509,Braun:1989qcd,CHERNYAK1990137,PhysRevD.44.3567} 
predictions for these transitions~\cite{Ball:2004rg, Khodjamirian:2006st, Bharucha:2012wy,Straub:2015ica},
which are typically applicable in the low $q^2$ regime, $q^2 \lesssim \mathcal{O}(m_b\lqcd) \sim 14$\,GeV.
In particular, we make use of the LCSR fit results for the BSZ parameters~\cite{Straub:2015ica}, comprising a fit to quadratic order in $z(q^2) - z(0)$. 
These results are shown in Table~\ref{tab:lcsr-coefficients}.
(For $b \to s$ transitions, Ref.~\cite{Straub:2015ica} also quotes combined fits of LCSR predictions with LQCD results, which are available for those decays.)

Importantly, the LCSR themselves generate correlated predictions between SM and NP form factors.
Thus, fitting these predictions in combination with measurements of the $q^2$ spectra for $B \to V \ell \nu$, in which only SM contributions are assumed,
nonetheless allows for predictions of improved precision for both the SM \emph{and} NP $B \to V$ form factors.
We proceed to perform such fits in Sec.~\ref{sec:theo_fit}.

\begin{table}
	\caption{LCSR prediction for the BSZ parameters in semileptonic $B\to \rho$ and $B\to \omega$ transitions. The full correlation matrices are given in~\cite{Straub:2015ica}.
	Note $\alpha_0^{A_0}$ and $\alpha_0^{T_2}$ are fixed with respect to $\alpha_0^{A_12}$ and $\alpha_0^{T_1}$, respectively, by the relations at $q^2=0$~\eqref{eqn:q20rel}.}
	\label{tab:lcsr-coefficients}
	\begin{tabular}{@{\extracolsep{0.75cm}}crr}
		\hline
		\hline
		Parameter           & $B \to \rho   \quad$   & $B \to \omega  \quad$   \\
		\hline
		$\alpha_1^{A_0}$    & $-0.83\pm0.20$              & $-0.83\pm0.30$                \\
		$\alpha_2^{A_0}$    & $1.33\pm1.05$               & $1.42\pm1.25$                 \\
		$\alpha_0^{A_1}$    & $0.26\pm0.03$               & $0.24\pm0.03$                 \\
		$\alpha_1^{A_1}$    & $0.39\pm0.14$               & $0.34\pm0.24$                 \\
		$\alpha_2^{A_1}$    & $0.16\pm0.41$               & $0.09\pm0.57$                 \\
		$\alpha_0^{A_{12}}$ & $0.30\pm0.03$               & $0.27\pm0.04$                 \\
		$\alpha_1^{A_{12}}$ & $0.76\pm0.20$               & $0.66\pm0.26$                 \\
		$\alpha_2^{A_{12}}$ & $0.46\pm0.76$               & $0.28\pm0.98$                 \\
		$\alpha_0^{V}$      & $0.33\pm0.03$               & $0.30\pm0.04$                 \\
		$\alpha_1^{V}$      & $-0.86\pm0.18$              & $-0.83\pm0.29$                \\
		$\alpha_2^{V}$      & $1.80\pm0.97$               & $1.72\pm1.24$                 \\
		$\alpha_0^{T_1}$    & $0.27\pm0.03$               & $0.25\pm0.03$                 \\
		$\alpha_1^{T_1}$    & $-0.74\pm0.14$              & $-0.72\pm0.22$                \\
		$\alpha_2^{T_1}$    & $1.45\pm0.77$               & $1.41\pm1.01$                 \\
		$\alpha_1^{T_2}$    & $0.47\pm0.13$               & $0.41\pm0.23$                 \\
		$\alpha_2^{T_2}$    & $0.58\pm0.46$               & $0.46\pm0.57$                 \\
		$\alpha_0^{T_{23}}$ & $0.75\pm0.08$               & $0.68\pm0.09$                 \\
		$\alpha_1^{T_{23}}$ & $1.90\pm0.43$               & $1.65\pm0.62$                 \\
		$\alpha_2^{T_{23}}$ & $2.93\pm1.81$               & $2.47\pm2.19$                 \\
		\hline
		\hline
	\end{tabular}
\end{table}

\section{Belle and \babar\ Spectrum Averages}\label{sec:average}
\label{sec:legacy-spectrum}
\begin{figure}
	\centering
	\includegraphics[width=\linewidth]{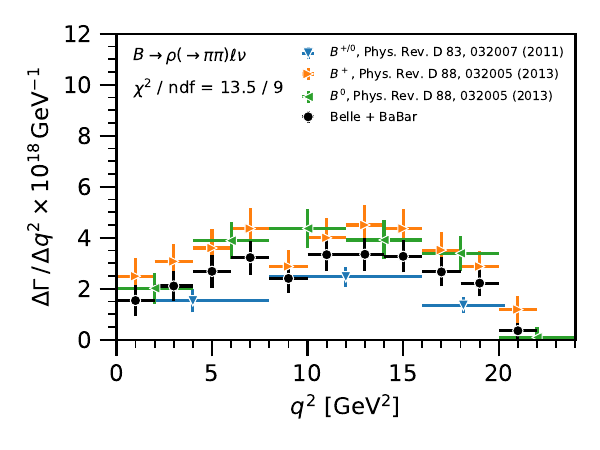}
	\includegraphics[width=\linewidth]{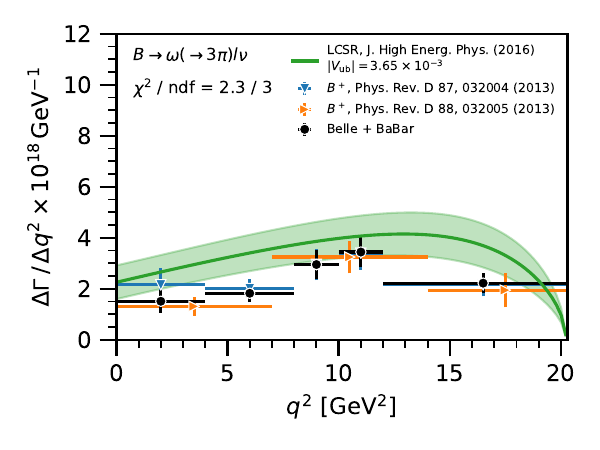}
	\caption{The averaged $q^2$ spectrum of the measurements listed in the text for the $\rho$ (top) and $\omega$ (bottom) final state on top of the latest Belle and \babar\ measurements. The isospin transformation is applied to the $B^0 \to \rho^-\ell^+\nu$ measurements. In the bottom figure we also show the model (green band) which was used to split the bins in the averaging procedure.}
	\label{fig:legacy-spectra}
\end{figure}

As a first step of our study, we generate averaged $q^2$ spectra from the measurements performed by the Belle and \babar\ experiments~\cite{Sibidanov:2013rkk,delAmoSanchez:2010af,Lees:2012mq}. 
To do this, we note the $B \to \rho \ell \bar\nu$ measurements of Belle and \babar\ have a compatible binning, 
which allows one to straightforwardly create an averaged differential spectrum. We define a $\chi^2$ function of the form
\begin{equation}
\begin{aligned}
	\chi^2(\bm{\bar{x}}) &= \kern-4.5em\sum_{\kern4em m \in \{\text{Belle, \babar}\}} \kern-4.5em \Delta \vec{y}^T_m \mat{C}^{-1}_m \Delta \vec{y}_m,\\
	\Delta \vec{y}_m &= \begin{pmatrix} 
	\vdots \\
	x_i^m - \kern-1.5em\sum\limits_{ \kern1em j > N_{i-1}}^{N_i} \kern-1em \bar{x}_j \\
	\vdots
	\end{pmatrix}\,,
\end{aligned}
\end{equation}
where $\mat{C}_m$ is the covariance of the measurement and $x_i^m$ is the measured differential rate in bin $i$ multiplied by the corresponding bin width. 
Further, $\bm{\bar{x}}$ denotes the averaged spectrum and $(N_{i-1},N_i]$ the range of averaged bins used to map to the $i$th measured bin. 
The binning of the averaged spectrum is chosen to match the most granular spectrum. 
The averaged spectrum is shown in black in Fig.~\ref{fig:legacy-spectra} and tabulated in Table.~\ref{tab:legacy-spectra}.

For the average of the $B \to \omega \ell \bar\nu$ measurements from Belle and \babar\, we again chose the binning of the most granular spectrum, in this case \babar's.
However the experimental spectra do not have a compatible binning in terms of matching bin boundaries.
In order to incorporate the Belle data and create an averaged spectrum, 
the LCSR fit results~\cite{Straub:2015ica} are used to create a model with which to split the second and fifth bin of the chosen binning, shown in black in Fig.~\ref{fig:legacy-spectra}.
To match the average bin onto a measurement without matching bin edges, the average bin $\bar{x}_i$, $i = 2$ or $5$, is split into two parts delimited by the lower bin edge, 
the $q^2$ value where the bin is split, and the upper bin edge. 
We label the two parts of the split bin as `left' and `right', respectively, in the following and define:
\begin{equation}
\begin{aligned}
	\bm{\bar{x}}_{i,\mathrm{left}} &= I_{i,\mathrm{left}}/I_i(1+\theta_i \varepsilon_{i,\mathrm{left}})\,, \\
	\bm{\bar{x}}_{i,\mathrm{right}} &= I_{i,\mathrm{right}}/I_i(1-\theta_i \varepsilon_{i,\mathrm{right}})\,,
\end{aligned}
\end{equation}
where $I_{i,\mathrm{left}}$ ($I_{i,\mathrm{right}}$) is the integral of the model function on the support of the left (right) part of the split bin, 
the sum $I_i = I_{i,\mathrm{left}} + I_{i,\mathrm{right}}$ is the integral over the entire bin, 
$\varepsilon_{i,\mathrm{left}}$ ($\varepsilon_{i,\mathrm{right}}$) the uncertainty of the integration given by the model uncertainty, 
and $\theta_i$ the nuisance parameter for the model dependence. 
We point out that the averaged spectrum does not depend on $|\Vub|$, as $|\Vub|$ cancels in the ratios $I_{i,\mathrm{left}} / I_i$ ($I_{i,\mathrm{right}} / I_i$).
The averaged spectrum is shown in black in Fig.~\ref{fig:legacy-spectra} and tabulated in Table~\ref{tab:legacy-spectra}. 
The result is almost independent of the nuisance parameters, as can be seen in the correlation matrix of the fit.

\begin{table}[tb]
	\caption{Averaged spectra. For the corresponding correlation matrices see Tables~\ref{tab:legacy-spectrum-correlation-rho} and~\ref{tab:legacy-spectrum-correlation-omega} in the Appendix.}
	\label{tab:legacy-spectra}
	\begin{tabular}{cc}
		\hline
		\hline
		\multicolumn{2}{c}{$B \rightarrow \rho \ell \bar \nu$} \\
		$q^2$ bin & $\Delta \Gamma / \Delta q^2 \times 10^6$ \\
		\hline
$[0,\,2]$ & $1.54 \pm 0.62$ \\
$[2,\,4]$ & $2.11 \pm 0.60$ \\
$[4,\,6]$ & $2.68 \pm 0.65$ \\
$[6,\,8]$ & $3.22 \pm 0.67$ \\
$[8,\,10]$ & $2.40 \pm 0.56$ \\
$[10,\,12]$ & $3.34 \pm 0.65$ \\
$[12,\,14]$ & $3.35 \pm 0.65$ \\
$[14,\,16]$ & $3.27 \pm 0.63$ \\
$[16,\,18]$ & $2.66 \pm 0.57$ \\
$[18,\,20]$ & $2.22 \pm 0.52$ \\
$[20,\,22]$ & $0.35 \pm 0.32$ \\
		\hline
		\hline
	\end{tabular}
	\hspace{1em}
	\begin{tabular}{cc}
		\hline
		\hline
		\multicolumn{2}{c}{$B \rightarrow \omega \ell \bar \nu$} \\
		$q^2$ bin & $\Delta \Gamma / \Delta q^2 \times 10^6$ \\
		\hline
$[0,\,4]$ & $1.51 \pm 0.46$ \\
$[4,\,8]$ & $1.82 \pm 0.35$ \\
$[8,\,10]$ & $2.95 \pm 0.56$ \\
$[10,\,12]$ & $3.44 \pm 0.59$ \\
$[12,\,21]$ & $2.22 \pm 0.40$ \\
		\hline
		\multicolumn{2}{c}{Nuisance Parameters} \\
		\hline
$\theta_2$ & $-0.01 \pm 1.00$ \\
$\theta_5$ & $0.00 \pm 1.00$ \\
		\hline
		\hline
		\vspace{2.76em}
	\end{tabular}
\end{table}

\begin{table}
	\caption{Averaged total branching ratio with $\tau_{B^+} = 1.638 \times 10^{-12}\,\mathrm{s}$ and the total rate given by the sum over the bins of the averaged spectra. The discrepancy between our result and the PDG arises because of the method of averaging. The PDG averages the directly measured branching ratios, whereas we average the provided unfolded spectra.}
	\begin{tabular}{ccc}
		\hline
		\hline
		Decay & \multicolumn{2}{c}{Branching Ratio} [$\times 10^{-4}$] \\
		& Our Result & PDG \\
		\hline
		$B^+ \to \rho^0 \ell \nu$ & $1.35 \pm 0.12$  & $1.58 \pm 0.11$\\
		$B^+ \to \omega \ell \nu$ & $1.14 \pm 0.13$ & $1.19 \pm 0.09$\\
		\hline
		\hline
	\end{tabular}
\end{table}

\section{Combined Data and Theory Fit}\label{sec:theo_fit}
\begin{figure}
	\centering
	\includegraphics[width=\linewidth]{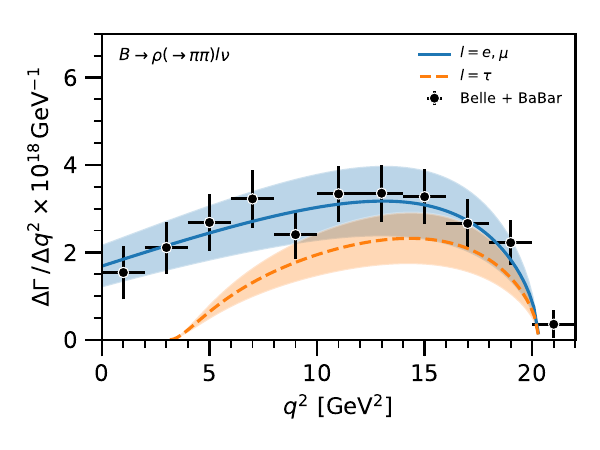}\\
	\includegraphics[width=\linewidth]{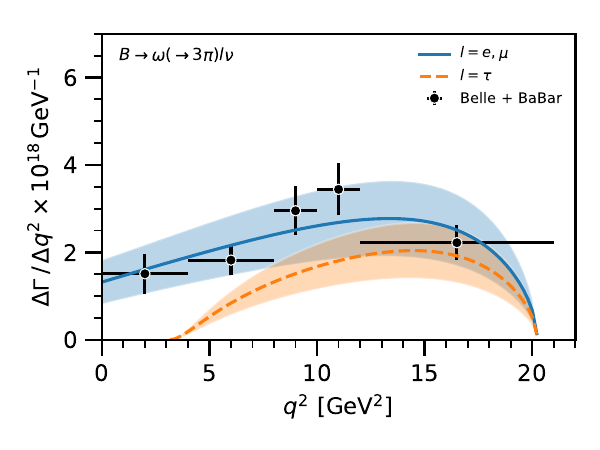}
	\caption{The differential decay rates for the leptonic and tauonic mode with our fit result for for the BSZ coefficients for $\BtoRhoLNu$ (top) and $\BtoOmegaLNu$ (bottom).}
	\label{fig:fit-result}
\end{figure}

We now fit the LCSR results in Table~\ref{tab:lcsr-coefficients} combined with the averaged spectra in Sec.~\ref{sec:legacy-spectrum} over the whole $q^2$ region,
thereby generating new predictions for the BSZ parameters beyond the $q^2 \lesssim 14\, \mathrm{GeV}^2$ regime of validity for the LCSR results.
To this end, we define a $\chi^2$ function of the form
\begin{equation}
\begin{aligned}
	\chi^2(|\Vub|, \vec{c}) &=  \Delta \vec{c}^T \mat{C}^{-1}_\mathrm{LCSR} \Delta \vec{c} + \Delta \vec{y}^T \mat{C}^{-1}_\mathrm{Spectrum} \Delta \vec{y} \, ,\\
	\Delta \vec{c} &= \vec{c}_\mathrm{LCSR} - \vec{c} \, , \\
	\Delta \vec{y} &= \vec{y}_\mathrm{Spectrum} - \vec{\Delta} \Gamma(\Vub, \vec{c}) / \vec{\Delta q^2} \, .
\end{aligned}
\end{equation}
Here, $\vec{c}$ denotes the vector of BSZ parameters and $\vec{y}$ is the binned differential decay rate. Note that $|\Vub|$ is included in the $\chi^2$ function and fitted simultaneously with the BSZ expansion coefficients.
We minimize the $\chi^2$ function using sequential least squares programming: 
The result of the fit is tabulated in Tab.~\ref{tab:fit-coefficients}, Tab.~\ref{tab:fit-correlation-rho}, Tab.~\ref{tab:fit-correlation-omega}. 
The differential rates for the leptonic and tauonic mode for both decays using our fitted coefficients are shown in Fig.~\ref{fig:fit-result}.  

We perform several cross-checks of our final fit. 
First, instead of a combined fit using the averaged spectrum described in Sec.~\ref{sec:legacy-spectrum}, 
we performed the fit with the individual spectra provided by the experiments. 
Second, the impact of the tensor form factors via their correlations to the non-tensor form factors was studied 
by fitting only the (SM) parameters contributing to the light lepton final state. 
Third, we sampled the form factors at different $q^2=0, 7, 14\,\mathrm{GeV}^2$ from an multi-dimensional Gaussian distribution, 
with mean and covariance set by the LCSR results, and incorporated these into the $\chi^2$ function. 
For each of these three cross-checks, no significant differences with respect to our combined fit results were found. 
This provides good evidence that our treatment of the form factors in the fit does not bias the result. 
The fit results for both final states are shown in Fig.~\ref{fig:fit-result}.

\begin{table}
	\caption{Fit result for $|\Vub|$ and the BCL expansion coefficients. The corresponding correlation matrices can be found in the appendix in 
	Tables~\ref{tab:fit-correlation-rho} and~\ref{tab:fit-correlation-omega}.}
	\label{tab:fit-coefficients}
	\begin{tabular}{@{\extracolsep{0.75cm}}crr}
		\hline
		\hline
		Parameter           & $B \to \rho  \quad $   & $B \to \omega  \quad $   \\
		\hline
		$|\Vub|$   	        & $2.96\pm0.29$               & $2.99\pm0.35$                 \\
		$\alpha_1^{A_0}$    & $-0.86\pm0.19$              & $-0.94\pm0.28$                \\
		$\alpha_2^{A_0}$    & $1.43\pm1.02$               & $1.78\pm1.20$                 \\
		$\alpha_0^{A_1}$    & $0.26\pm0.03$               & $0.24\pm0.03$                 \\
		$\alpha_1^{A_1}$    & $0.38\pm0.13$               & $0.30\pm0.22$                 \\
		$\alpha_2^{A_1}$    & $0.16\pm0.41$               & $0.00\pm0.55$                 \\
		$\alpha_0^{A_{12}}$ & $0.29\pm0.03$               & $0.25\pm0.04$                 \\
		$\alpha_1^{A_{12}}$ & $0.72\pm0.17$               & $0.54\pm0.24$                 \\
		$\alpha_2^{A_{12}}$ & $0.37\pm0.70$               & $-0.03\pm0.96$                \\
		$\alpha_0^{V}$      & $0.33\pm0.03$               & $0.31\pm0.04$                 \\
		$\alpha_1^{V}$      & $-0.87\pm0.18$              & $-0.89\pm0.27$                \\
		$\alpha_2^{V}$      & $1.88\pm0.94$               & $1.81\pm1.19$                 \\
		$\alpha_0^{T_1}$    & $0.27\pm0.03$               & $0.25\pm0.03$                 \\
		$\alpha_1^{T_1}$    & $-0.75\pm0.14$              & $-0.76\pm0.21$                \\
		$\alpha_2^{T_1}$    & $1.51\pm0.76$               & $1.50\pm0.96$                 \\
		$\alpha_1^{T_2}$    & $0.46\pm0.13$               & $0.37\pm0.21$                 \\
		$\alpha_2^{T_2}$    & $0.59\pm0.46$               & $0.38\pm0.55$                 \\
		$\alpha_0^{T_{23}}$ & $0.74\pm0.07$               & $0.65\pm0.09$                 \\
		$\alpha_1^{T_{23}}$ & $1.83\pm0.40$               & $1.40\pm0.58$                 \\
		$\alpha_2^{T_{23}}$ & $2.88\pm1.79$               & $2.03\pm2.18$                 \\
		\hline
		\hline
	\end{tabular}
\end{table}

\begin{figure}
	\centering
	\includegraphics[width=\linewidth]{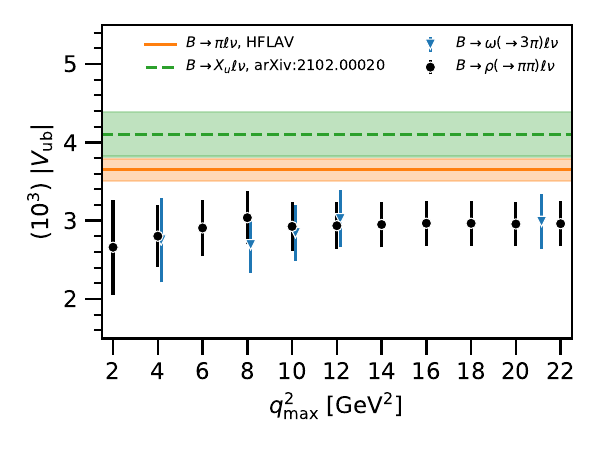}
	\caption{The extracted $|\Vub|$ values from $\BtoRhoLNu$ and $\BtoOmegaLNu$ for different cut-offs $q^2_{\mathrm{max}}$ of the respective $q^2$ spectrum in the fit. The stable extraction of $\Vub$ for increasing $q^2$ cut-offs indicates that the extrapolation into the high $q^2$ region works.}
	\label{fig:vub-summary}
\end{figure}

\begin{figure}
	\centering
	\includegraphics[width=\linewidth]{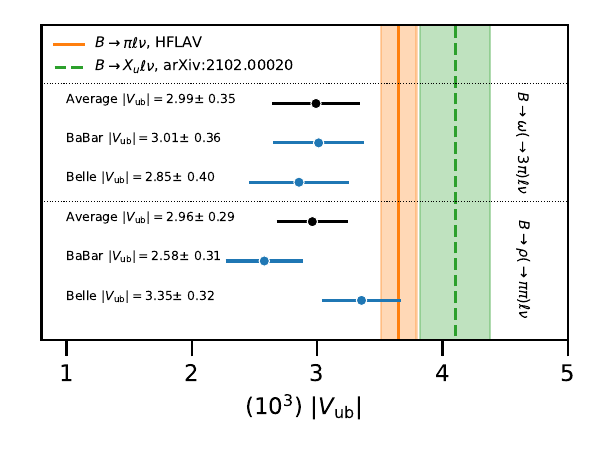}
	\caption{The extracted $|\Vub|$ values from $\BtoRhoLNu$ and $\BtoOmegaLNu$ for the fits to the individual experiments, and our averaged spectra. The $\BtoRhoLNu$ measurements of Belle and \babar exhibit a slight tension.}
	\label{fig:vub-experiments}
\end{figure}

We find that the extracted $|\Vub|$ is consistently smaller in comparison to the extraction from $\BtoPiLNu$ decays.
Our extracted values for $|\Vub|$ are compatible with the extractions in Ref.~\cite{Straub:2015ica}, 
but yield lower uncertainties, because we extract $|\Vub|$ from a combination of experiments and LCSR results over the full $q^2$ range 
instead of individually for the Belle and \babar experiments with different $q^2_\mathrm{max}$ cut-offs. 
As a cross-check, we have repeated the fit with different cut-offs of the measured $q^2$ spectrum. 
The results of these fits are shown in Fig.~\ref{fig:vub-summary}. 
We consistently find central values for $|\Vub|$ from $\BtoRhoLNu$, and $\BtoOmegaLNu$ below the value for $|\Vub|$ extracted from $\BtoPiLNu$. 
The stable $|\Vub|$ extraction for increasing $q^2$ cut-offs indicates that the extrapolation of the form factors into the high $q^2$ region is reliable.
We also perform the fits to extract $|\Vub|$ for each experiment separately due to the large discrepancy in the measured $\BtoRhoLNu$ spectra between Belle and \babar. 
The results of these individual fits is summarized in Fig.~\ref{fig:vub-experiments}. We find that for the $\rho$ channel, the measurements of Belle and \babar exhibit a slight tension.

\section{Predictions in the Standard Model and Beyond}\label{sec:preds}
Using our combined fit, in Table~\ref{tab:observables} we provide SM predictions for the lepton universality ratios $\RRho$ and $\ROmega$,
defined as usual as
\begin{equation}
 	\RV = \frac{\Gamma(\BtoVTauNu)}{\Gamma(B \to V \ell \bar \nu)}\,.
\end{equation}
The combined fit improves the prediction for these observables over using the LCSR fit results alone by $24$\% and $13$\%, respectively.
It is further interesting to consider phase space constrained lepton universality ratios, as pointed out by Refs.~\cite{Freytsis:2015qca,Bernlochner:2016bci}, 
\begin{equation}
	\redRV = \frac{\int_{m_\tau^2}^{t_-}dq^2\, [d\Gamma(\BtoVTauNu)/dq^2]}{\int_{m_\tau^2}^{t_-} dq^2\,[d\Gamma(\BtoVLNu)/dq^2]} \, ,
\end{equation}
i.e. restricting the light lepton mode to $m_{\tau}^2 \le q^2 \le (m_B - m_V)^2 \equiv t_-$, such that the phase space suppression of the $\tau$ mode is lifted. 
In $\redRV$, the correlation is increased between nominator and denominator and thus a larger cancellation of uncertainties is possible, 
but a small dependence on the actual shape of the light-lepton differential rate is introduced by the cut-off at $m_\tau^2$.
$\redRV$ is insensitive to the low $q^2 \le m_\tau^2 \simeq 3.16$\,GeV$^2$ regime, 
reducing its sensitivity to data in the nominal regime of validity of the light-cone expansion $q^2 \lesssim 14$\,GeV$^2$.
However, we see in Table~\ref{tab:observables} that the LCSR predictions for $\redRRho$ and $\redROmega$ are in good agreement with the combined fit,
suggesting that the experimental data does not pull the (extrapolation of the) LCSR fit results significantly in the higher $q^2$ regime.

We also calculate SM predictions for several angular observables, utilizing our combined fit result for the form factors.
First, we consider the vector meson longitudinal polarization fraction
\begin{equation}
	\label{eqn:FLrho}
	F_{L,l}(V) = \frac{\Gamma_{\lambda=0}(\BtoVLNu)}{\Gamma(\BtoVLNu)}\,, \\
\end{equation}
with $\lambda$ the helicity of the vector meson $V = \rho$, $\omega$. 
As an aside, in the $B \to (\rho \to \pi\pi)l \bar\nu$ decay, it is well-known that the longitudinal polarization of the $\rho$ arises 
in the differential rate with respect to the pion polar helicity angle, as in Eq.~\eqref{eqn:FLrhodiff}.
One may derive a similar result for the $\omega$ longitudinal polarization in $B \to (\omega \to \pi\pi\pi) l \bar\nu$, 
via the Dalitz-type analysis provided in App.~\ref{sec:amplrates}, yielding
\begin{equation}
	\label{eqn:FLomegadiff}
	\frac{1}{\Gamma} \frac{d \Gamma}{d \cos\theta_+} = \frac{3}{8}\bigg[\!\big[1 - F_{L}(\omega)\big](1 + \cos^2\theta_+) + 2F_{L}(\omega) \sin^2\theta_+\!\bigg]\,,
\end{equation}
in which the $\theta_+$ helicity angle defines the angle between the $\pi^+$ momentum and the $B$ momentum $\bm{p}_{B}$ in the $\omega$ rest frame.
Second, we calculate the $\tau$ polarization (see e.g.\,\cite{Bernlochner:2021vlv})
\begin{equation}
	P_{\tau}(V) = \frac{\Gamma_{+}(\BtoVTauNu) - \Gamma_{-}(\BtoVTauNu)}{\Gamma_{+}(\BtoVTauNu) + \Gamma_{-}(\BtoVTauNu)}\,,
\end{equation}
in which the $\pm$ subscript labels the $\tau$ helicity, such that one obtains $P_{\tau}(V) = -1$ in the SM for $m_\tau \to 0$. We also calculate the forward-backward asymmetry
\begin{equation}
	A_{\mathrm{FB},l}(V)  = \frac{\Gamma_{[0,1]}(B \to V l \bar\nu) - \Gamma_{[-1,0]}(B \to V l \bar\nu)}{\Gamma_{[0,1]}(B \to V l \bar\nu) + \Gamma_{[-1,0]}(B \to V l \bar\nu)}\,,
\end{equation}
in which $\Gamma_L = \int_L d\!\cos\theta_l \, [d\Gamma/d\!\cos\theta_l]$.
The predicted central values and uncertainties for these observables are shown in Table~\ref{tab:observables}. 
Using the fitted form factors improves the prediction for these angular observables over using the LCSR fit results alone by up to $21$\%.

\begin{table}
	\caption{Predictions for the tauonic to leptonic ratios $\RV$, $\redRV$, 
	the longitudinal fractions $F_L$, the $\tau$ polarization, 
	and the forward-backward asymmetries using the LCSR predictions and our combined fit results for the BSZ parameters.}
	\label{tab:observables}
	\centering
	\begin{tabular}{lrrc}
		\hline
		\hline
		 & \multicolumn{1}{c}{\quad LCSR~\cite{Straub:2015ica}} & \multicolumn{1}{c}{Fit} & Improvement \\
		\hline
$\RRho$      & $0.532 \pm 0.011$ & $0.535 \pm 0.009$ & $24\,\%$           \\
$\redRRho$   & $0.605 \pm 0.007$ & $0.606 \pm 0.007$ & $4\,\%$ \\				
$F_{L,\ell}(\rho)$ & $0.512 \pm 0.068$ & $0.498 \pm 0.058$ & $15\,\%$ \\
$F_{L,\tau}(\rho)$ & $0.496 \pm 0.062$ & $0.482 \pm 0.052$ & $16\,\%$ \\
$P_\tau(\rho)$ & $-0.543 \pm 0.025$ & $-0.552 \pm 0.020$ & $21\,\%$ \\
$A_{\mathrm{FB},\ell}(\rho)$ & $-6.641 \pm 0.769$ & $-6.773 \pm 0.644$ & $16\,\%$ \\
$A_{\mathrm{FB},\tau}(\rho)$ & $-2.023 \pm 0.705$ & $-2.214 \pm 0.615$ & $13\,\%$ \\ 	
		\hline
$\ROmega$    & $0.534 \pm 0.018$ & $0.543 \pm 0.015$ & $13\,\%$           \\
$\redROmega$ & $0.606 \pm 0.012$ & $0.610 \pm 0.011$ & $5\,\%$ \\		
$F_{L,\ell}(\omega)$ & $0.501 \pm 0.071$ & $0.472 \pm 0.067$ & $6\,\%$ \\
$F_{L,\tau}(\omega)$ & $0.486 \pm 0.069$ & $0.465 \pm 0.065$ & $5\,\%$ \\
$P_\tau(\omega)$ & $-0.545 \pm 0.029$ & $-0.554 \pm 0.028$ & $2\,\%$ \\
$A_{\mathrm{FB},\ell}(\omega)$ & $-6.604 \pm 0.868$ & $-7.015 \pm 0.852$ & $2\,\%$ \\
$A_{\mathrm{FB},\tau}(\omega)$ & $-2.102 \pm 0.849$ & $-2.455 \pm 0.834$ & $2\,\%$ \\
		\hline
		\hline		
	\end{tabular}
\end{table}

\begin{figure}
	\centering
	\includegraphics[width=\linewidth]{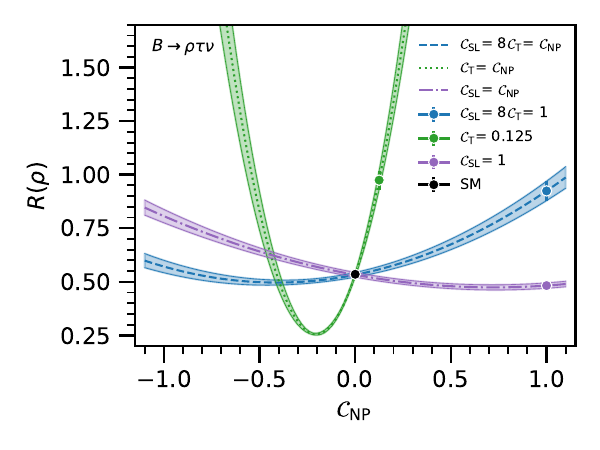}
	\caption{The impact on the lepton universality ratio $\RRho$ for the leptoquark model $R_2$, with Wilson coefficients $c_{\SL} \simeq 8 c_T$. 
	Additionally, the individual contributing NP currents to $R_2$ are shown. The highlighted NP points correspond to the benchmark points for Fig.~\ref{fig:mm2-scalar-np}.}
	\label{fig:cnp-rrho-np}
\end{figure}

We may further use our combined fit to examine the effects of NP operators, defined in Eq.~\eqref{eqn:LeffNP}, on $B \to V \tau \bar\nu$ decays. 
(These effects are the same for either $\rho$ or $\omega$---both are vector mesons---up 
to small differences from their slightly different masses and their disparate decay modes.)
As an example, in Fig.~\ref{fig:cnp-rrho-np} we show the variation in $\RRho$ for the leptoquark simplified model $R_2$ \cite{Dorsner:2016wpm,BUCHMULLER1987442}.
In this model, a heavy TeV-scale leptoquark mediator induces non-zero $c_{\SL}$ and $c_T$ NP Wilson coefficients, 
constrained such that $c_{\SL} \simeq 8 c_T$ once Fierz relations and RG evolution effects are included.
Over the range of NP couplings considered, $\RRho$ varies by almost a factor of two.

\begin{figure}
	\centering
	\includegraphics[width=\linewidth]{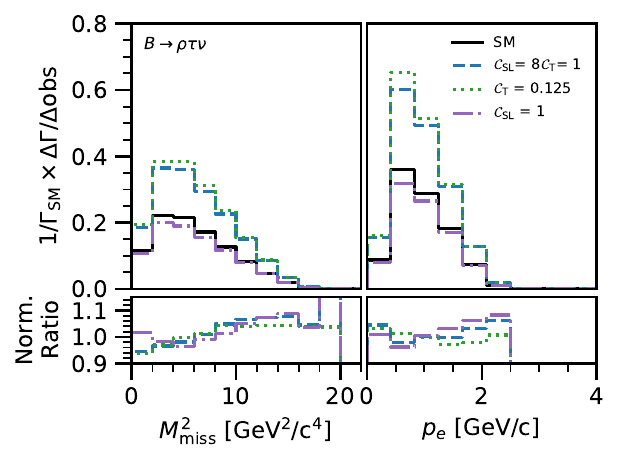}
	\caption{The $B \to \rho \tau \bar \nu_\tau$ distribution in the missing mass squared $M_\mathrm{miss}^2$ variable (left) and the lepton momentum in the $B$ rest frame (right) without reconstruction effects, 
	for the $R_2$ leptoquark model benchmark point $c_{\SL} = 8 c_T = 1$, and the individual currents at the benchmark points $c_{\SL} = 1$ and $c_T = 0.125$. 
	The differential distributions are normalized with respect to the SM rate and include a cut on the electron momentum in the lab frame $p_e > 300\,\mathrm{MeV}$.
	In the lower panels, we show the ratio of the shapes of differential distributions, with all distributions normalized to unity.}
	\label{fig:mm2-scalar-np}
\end{figure}

In Fig.~\ref{fig:mm2-scalar-np} for the benchmark choice $c_{\SL} = 8c_T = 1$, we show the effects on the differential distributions 
in missing mass squared $M_\mathrm{miss}^2$ and the electron momentum $p_e$ from the $\tau$ decay compared to the SM.
These spectra are generated using the \texttt{Hammer} library~\cite{Bernlochner:2020tfi,bernlochner_florian_urs_2020_3993770}
to reweight a sample of $5\times10^4$ events, generated with \texttt{EvtGen R01-07-00}~\cite{Lange:2001uf}.
We have imposed a common experimental threshold that the lepton momentum be greater than $300\,\mathrm{MeV}$, 
but otherwise we do not consider reconstruction effects.
At the benchmark point, the $R_2$ couplings generate deviations of approximately $5$--$10\%$ compared with the SM distributions.
Just as for analyses of $b \to c \tau \bar\nu$ decays, using the full differential information is expected to provide greater sensitivity to NP effects than considering deviations in $\RV$ (or $R(\pi)$) alone.
Moreover, once high precision measurements for these decays are available, 
self-consistent analyses, using e.g. reweighting tools, may be required to avoid biases in NP interpretations of future anomalous $\RV$ measurements (if any)~\cite{Bernlochner:2020tfi}.

\begin{figure}
	\centering
	\includegraphics[width=\linewidth]{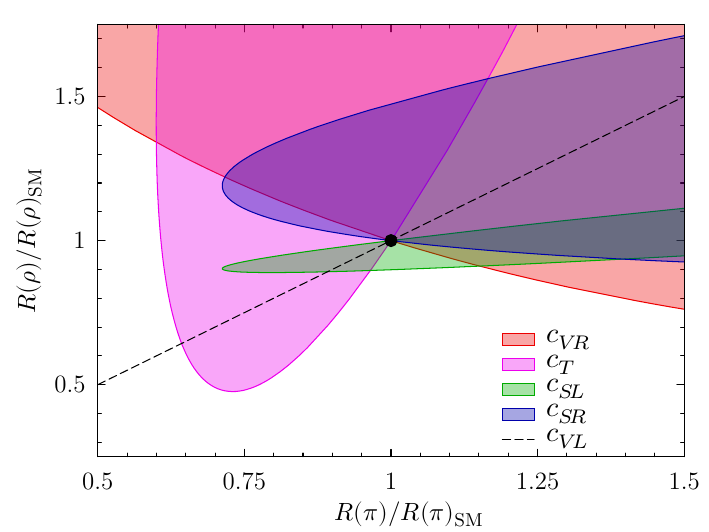}
	\caption{Allowed regions in the $\RRho/\RRho_{\text{SM}}$ versus $R(\pi)/R(\pi)_{\text{SM}}$ plane,
	 for each of the (complex) NP couplings $c_{\SR}$, $c_{\SL}$, $c_{\VR}$ and $c_T$ combined with the SM contribution.
	The coupling $c_{\VL}$ simply rescales the SM, and therefore spans only a straight line contour.}
	\label{fig:rp-rv}
\end{figure}

Finally, it is perhaps also instructive to characterize the interplay between $\RV$ and $R(\pi)$:
Unlike for $B \to D^{(*)}$ there are no heavy quark symmetry relations between the vector and pseudoscalar meson decay modes.
To this end, in Fig.~\ref{fig:rp-rv} we show the allowed regions in the $\RRho$--$R(\pi)$ plane, 
for each of the (complex) couplings $c_{\SR}$, $c_{\SL}$, $c_{\VR}$ and $c_T$.
For the $R(\pi)$ NP predictions, we use the LCSR fit of Ref.~\cite{Gubernari:2018wyi}.
However, we note the SM prediction therefrom is $R(\pi)_{\text{SM}}  = 0.75\pm0.02$, 
which is quite different to the SM prediction from the combination of LQCD calculations and experimental data $R(\pi)_{\text{SM}} = 0.641 \pm 0.016$~\cite{Bernlochner:2015mya}. 
(A more recent analysis using LCSR inputs yields $0.688 \pm 0.014$~\cite{Leljak:2021vte}, which is still in some tension.)
For this reason, in Fig.~\ref{fig:mm2-scalar-np} we plot $\RRho/\RRho_{\text{SM}}$ and $R(\pi)/R(\pi)_{\text{SM}}$, 
assuming that any LCSR pulls on $R(\pi)$ approximately factor out of these normalized ratios.
The allowed regions for each NP coupling are broadly similar to those in the $R(D)$--$R(D^*)$ plane (see e.g. Refs.~\cite{Tanaka:2012nw, Ligeti:2016npd}).

\section{Summary and Conclusions}
Using our generated averages of the $\BtoRhoLNu$ and $\BtoOmegaLNu$ differential spectra measured by the Belle and \babar experiments, 
we performed a combined fit with LCSR results to obtain improved predictions for the $B \to V$ form factors, for $V = \rho$ or $\omega$, over the full $q^2$ range.

With our combined fit results, we extracted $|\Vub|$ from the averaged spectra in both decay modes,
\begin{equation}
	\begin{aligned}
		|\Vub|_{\BtoRhoLNu} &= (2.96 \pm 0.29) \times 10^{-3} \,,\\
		|\Vub|_{\BtoOmegaLNu} &= (2.99 \pm 0.35) \times 10^{-3}\,, \\
	\end{aligned}
\end{equation}
finding $|\Vub|$ consistently below other inclusive and exclusive extractions and with smaller uncertainty compared to previous extractions such as in~\cite{Straub:2015ica}.
We further used our combined fit to calculate the following set of observables in the SM: 
The lepton universality rations $\RV$, the longitudinal polarization fractions $F_{L,l}(V)$, 
the $\tau$ polarization $P_\tau(V)$, and the forward-backward asymmetry $A_{\mathrm{FB},l}(V)$. 
For these observables, we see improved precision in the predictions, by up to 24\,\% compared to using the LCSR results alone.
In addition, we briefly investigated the impact of New Physics contributions on $\RV$ in the $B\to V$ transitions for all four-Fermi NP operators, 
as well as examining the impacts on differential rates for a benchmark example using the leptoquark model $R_2$.

We look forward to future lattice QCD predictions near zero recoil and beyond, which can provide additional constraints on this combined fit in the high $q^2$ regime. 
We also look forward to new measurements of differential spectra for $\BtoVLNu$ from Belle~II and LHCb. 
These measurements might help to resolve the tension seen in the $\BtoRhoLNu$ spectra from Belle and \babar, 
and to investigate the consistently smaller values of $|\Vub|$ extracted from both channels.

\begin{acknowledgments}

We thank Aoife Bharucha, David Straub, Roman Zwicky for helpful discussions. 
FB is supported by DFG Emmy-Noether Grant No. BE 6075/1-1 and BMBF Grant No. 05H19PDKB1.
MP is supported by the Argelander Starter-Kit Grant of the University of Bonn and BMBF Grant No. 05H19PDKB1.
DJR is supported in part by the Office of High Energy Physics of the U.S. Department of Energy under contract DE-AC02-05CH11231.

\end{acknowledgments}

\appendix

\section{Amplitudes and Differential Rates}
\label{sec:amplrates}

We write explicit expressions for the $\bbar \to \ubar$ amplitudes rather than $b \to u$, defining the basis of NP operators to be
\begin{subequations}\label{abdef}
\begin{align}
\text{SM:\,} & \phantom{-}2\sqrt{2}\, \Vub^* G_F\big[\bbar \g^\mu P_L u\big] \big[\bar\nu \g_\mu P_L l\big]\,, \\*
\text{Vector:\,} & \phantom{-}2\sqrt{2}\, \Vub^* G_F
  \big[\bbar\big(\alVL \g^\mu P_L + \alVR \g^\mu P_R\big)u\big] \nn \\
  &\times\big[\bar\nu\big(\beVL \g_\mu P_L + \beVR \g_\mu P_R\big) l\big]\,, \\
\text{Scalar:\,} & -2\sqrt{2}\, \Vub^* G_F
  \big[\bbar\big(\alSL P_L + \alSR P_R\big)u\big] \nn \\
  &\times \big[\bar\nu\big(\beSL P_R + \beSR P_L\big) l\big]\,, \\
\text{Tensor:\,} & -2\sqrt{2}\, \Vub^* G_F
  \big[ \big(\bbar \alTR \sigma^\mn P_R u\big)
  \big(\bar\nu \beTL \sigma_\mn P_R l \big)  \nn\\
  & + \big(\bbar \alTL \sigma^\mn P_L u\big)
  \big(\bar\nu \beTR \sigma_\mn P_L l \big) \big]\,,
\end{align}
\end{subequations}
with $l = e$, $\mu$, $\tau$.
The subscript of the $\beta$ coupling denotes the $\nu$ chirality and the subscript of the $\alpha$ coupling is that of the $u$ quark.
Operators for the CP conjugate $b \to u$ processes follow by Hermitian conjugation.
The correspondence between the $\alpha$, $\beta$ coefficients and the basis typically chosen for, 
e.g., $b\to c$ or $b\to u$ operators can be found in Ref.~\cite{Bernlochner:2017jxt}.
With respect to the notation in Eq.~\eqref{eqn:LeffNP},
\begin{align}
	c_{\SR}^* & = -\alSL \beSL\,, & c_{\SL}^* & = -\alSR \beSL\,, \nn \\
	c_{\VR}^* & = \alVR \beVL \,, & c_{\VL}^* & = \alVR \beVL\,, \nn \\
	c_T^*  & = -\alTR\beTL\,.
\end{align}

\begin{figure}[t]
	\includegraphics[width = 0.4\linewidth]{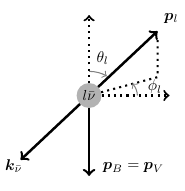}
	\includegraphics[width = 0.4\linewidth]{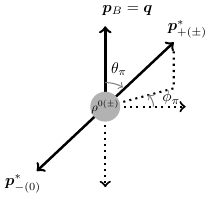}
	\includegraphics[width = 0.4\linewidth]{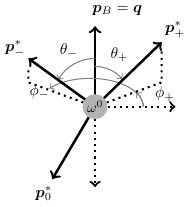}
	\caption{Definition of the helicity angles in the lepton system, $\rho$, and $\omega$ rest frames.
	In the case of the $\rho^0 \to \pi^+ \pi^-$ ($\rho^\pm \to \pi^\pm \pi^0$) decay, the helicity angles $\theta_\pi$ and $\phi_\pi$ are defined with respect to the $\pi^+$ ($\pi^\pm$).}
	\label{fig:hadef}
\end{figure}

The $\BtoVLNu$ decay has three external quantum numbers: 
$\lambda_V = \pm,0$,   $s_l = 1,2$, and $s_\nu = \pm$, 
which are the vector meson and massive lepton spin and neutrino helicity, respectively.
(We label the $s_l$ spin by `$1$' and `$2$', rather than `$-$' and `$+$', to match the conventions of Ref.~\cite{Ligeti:2016npd} for massive spinors on internal lines.)
Helicity angles are similarly defined with respect to the $\bbar \to \ubar$ process;
definitions for the conjugate process follow simply by replacing all particles with their antiparticles. 
The azimuthal helicity angle $\phtau$ of the $\bm{p}_l$--$\bm{k}_{\bar\nu}$ plane, defined in the $l\bar\nu$ center-of-mass frame, is unphysical in the pure $B \to V l\bar\nu$ decay.
See Fig.~\ref{fig:hadef}.
The single physical polar helicity angle, $\thtau$, defines the orientation of $\bm{p}_l$ 
in the lepton center of mass reference frame, with respect to $-\bm{p}_{B}$.

For compact expression of the amplitudes, it is further convenient to define 
\begin{gather}
	\mSqq = q^2/m_{B}^2\,, \quad \rl = m_l/m_{B}\,,  \nn\\
	\quad \rU = m_V/m_B\,, \quad \mPw = |p_V|/m_B\,,
\end{gather}	
in which the spatial momentum $|p_V| = m_V \sqrt{w^2-1}$, is the momentum of the vector meson in the $B$ rest frame.
We remove an overall prefactor $2 c_V G_F \Vub^*  m_{B}^2 \sqrt{\mSqq - \rl^2}$ from the amplitudes, 
in which the coefficient $c_V = 1/\sqrt{2}$ for the neutral unflavored mesons final states $\rho^0$ and $\omega^0$, while $c_V = 1$ for $\rho^\pm$.
Thus, the $\BtoVLNu$ full differential rate
\begin{multline}
	\label{eqn:diffrate}
	\frac{d^2\Gamma}{dq^2 \, d\cos\thtau} = \frac{c_V^2 G_F^2 |\Vub|^2 m_{B}^3}{64\pi^3}\mPw\frac{(\mSqq -\rl^2)^2}{\mSqq} \\
	\times \sum_{\lambda_V,s_l, s_\nu}\!\!\!\big|A_{\lambda_V s_l s_\nu}\big|^2 \,,
\end{multline}
and the $\bbar \to \ubar l \nu$ amplitudes are correspondingly
\begin{subequations}
\label{eqn:buampls}
\begin{align}
A_{-\,1\,\dn} & = \sin\thtau\bigg\{-\frac{\FV (1+(\alVL+\alVR) \beVL) \mPw \rl}{\sqrt{\mSqq} (1+\rU)} \nn \\
 		& -\frac{\FA1 (1+(\alVL-\alVR) \beVL) \rl (1+\rU)}{2 \sqrt{\mSqq}} \nn \\
 		& +\frac{2 \alTR \beTL \big[\FT2+2 \FT1 \mPw-\FT2 \rU^2\big]}{\sqrt{\mSqq}}\bigg\}\\
A_{-\,2\,\dn} & = \cos^2 \frac{\thtau}{2}\bigg\{-\frac{2 \FV (1+(\alVL+\alVR) \beVL) \mPw}{1+\rU} \nn \\
 		& -\FA1 (1+(\alVL-\alVR) \beVL) (1+\rU) \nn \\
 		& +\frac{4 \alTR \beTL \rl \big[\FT2+2 \FT1 \mPw-\FT2 \rU^2\big]}{\mSqq}\bigg\}\\
A_{-\,1\,\up} & = \sin^2 \frac{\thtau}{2}\bigg\{\frac{2 \FV (\alVL+\alVR) \beVR \mPw}{1+\rU} \nn \\
 		& +\FA1 (\alVL-\alVR) \beVR (1+\rU) \nn \\
 		& +\frac{4 \alTL \beTR \rl \big[\FT2-2 \FT1 \mPw-\FT2 \rU^2\big]}{\mSqq}\bigg\}\\
A_{-\,2\,\up} & = \sin\thtau\bigg\{\frac{\FV (\alVL+\alVR) \beVR \mPw \rl}{\sqrt{\mSqq} (1+\rU)} \nn \\
 		& +\frac{\FA1 (\alVL-\alVR) \beVR \rl (1+\rU)}{2 \sqrt{\mSqq}} \nn \\
 		& +\frac{2 \alTL \beTR \big[\FT2-2 \FT1 \mPw-\FT2 \rU^2\big]}{\sqrt{\mSqq}}\bigg\}\\
A_{0\,1\,\dn} & = \bigg\{\frac{\AP (-\alSL+\alSR) \beSL \mPw}{\sqrt{2} \rU} \nn \\
 		& -\frac{\sqrt{2} (1+(\alVL-\alVR) \beVL) \rl (\FA0 \mPw-4 \FA{12} \rU \cos\thtau)}{\mSqq} \nn \\
 		& -\frac{8 \sqrt{2} \FT{23} \alTR \beTL \rU \cos\thtau}{1+\rU}\bigg\}\\
A_{0\,2\,\dn} & = \sin\thtau\bigg\{-\frac{4 \sqrt{2} \FA{12} (1+(\alVL-\alVR) \beVL) \rU}{\sqrt{\mSqq}} \nn \\
 		& +\frac{8 \sqrt{2} \FT{23} \alTR \beTL \rl \rU}{\sqrt{\mSqq} (1+\rU)}\bigg\}\\
A_{0\,1\,\up} & = \sin\thtau\bigg\{\frac{4 \sqrt{2} \FA{12} (-\alVL+\alVR) \beVR \rU}{\sqrt{\mSqq}} \nn \\
 		& -\frac{8 \sqrt{2} \FT{23} \alTL \beTR \rl \rU}{\sqrt{\mSqq} (1+\rU)}\bigg\}\\
A_{0\,2\,\up} & = \bigg\{\frac{\AP (\alSL-\alSR) \beSR \mPw}{\sqrt{2} \rU} \nn \\
 		& +\frac{\sqrt{2} (\alVL-\alVR) \beVR \rl (\FA0 \mPw-4 \FA{12} \rU \cos\thtau)}{\mSqq} \nn \\
 		& -\frac{8 \sqrt{2} \FT{23} \alTL \beTR \rU \cos\thtau}{1+\rU}\bigg\}\\
A_{+\,1\,\dn} & = \sin\thtau\bigg\{\frac{\FV (1+(\alVL+\alVR) \beVL) \mPw \rl}{\sqrt{\mSqq} (1+\rU)} \nn \\
 		& -\frac{\FA1 (1+(\alVL-\alVR) \beVL) \rl (1+\rU)}{2 \sqrt{\mSqq}} \nn \\
 		& +\frac{2 \alTR \beTL \big[\FT2-2 \FT1 \mPw-\FT2 \rU^2\big]}{\sqrt{\mSqq}}\bigg\}\\
A_{+\,2\,\dn} & = \sin^2 \frac{\thtau}{2}\bigg\{-\frac{2 \FV (1+(\alVL+\alVR) \beVL) \mPw}{1+\rU} \nn \\
 		& +\FA1 (1+(\alVL-\alVR) \beVL) (1+\rU) \nn \\
 		& +\frac{4 \alTR \beTL \rl \big[2 \FT1 \mPw+\FT2 \big[\rU^2 - 1\big]\big]}{\mSqq}\bigg\}\\
A_{+\,1\,\up} & = \cos^2 \frac{\thtau}{2}\bigg\{\frac{2 \FV (\alVL+\alVR) \beVR \mPw}{1+\rU} \nn \\
 		& +\FA1 (-\alVL+\alVR) \beVR (1+\rU) \nn \\
 		& -\frac{4 \alTL \beTR \rl \big[\FT2+2 \FT1 \mPw-\FT2 \rU^2\big]}{\mSqq}\bigg\}\,.
\end{align}
\end{subequations}

As done in Refs.~\cite{Ligeti:2016npd, Bernlochner:2017jxt, Bernlochner:2018kxh}, 
in Eqs.~\eqref{eqn:buampls} we have adopted spinor conventions such that unphysical $\phtau$ phase is removed from the $\bbar \to \ubar l \nu$ amplitude, 
transferring it to the subsequent $\tau$ or $V$ vector meson decays to generate physical phase combinations therein.
In particular, if subsequent  $V \to X_1\ldots X_n$ decays are included, one may further define helicity angles $\phi_{ij}$ with respect to the $X_i$--$X_j$ plane, 
such that the twist angle $\phtau - \phi_{ij}$ becomes a physical phase in the $V \to X_1\ldots X_n$ amplitude.
Similarly, $\tau \to h \nu$ decays, for $h$ any final state system,
feature a helicity angle $\phi_h$ defined by the $h$--$\nu$ plane, such that $\phtau-\phi_h$ becomes physical in the $\tau$ decay amplitude.
With respect to the explicit amplitudes $A_{\lambda_V s_l s_\nu}$ in Eqs.~\eqref{eqn:buampls}, 
this phase transference amounts to requiring the inclusion of an additional spinor phase function in the subsequent $\tau$ and $V$ decay amplitudes:
$h^l_{s_l s_{\nu}}$ and $h^V_{\lambda_V}$, respectively, 
that modify the usual phase convention of the $\tau$ or $V$ helicity basis.
These two functions are defined exhaustively via $h^l_{1 \dn} = h^{l*}_{2 \up} = 1$, $h^l_{1 \up} = h^{l *}_{2 \dn} = e^{i\phtau}$ and $h^V_{\lambda_V} = e^{-i\lambda_{V} \phtau}$.

To incorporate subsequent $\rho \to 2\pi$ or $\omega \to 3\pi$ decays, the full differential rate  can be written as
\begin{multline}
	d\Gamma = \frac{G_F^2 |\Vub|^2 c_V^2 m_{B}^3}{128\pi^4}\mPw\frac{(\mSqq -\rl^2)^2}{\mSqq} \\ \times\sum_{s_l, s_\nu}\!\!\big|A_{s_l s_\nu}\big|^2 dq^2 \,d\Omega_l \,d \mathcal{PS}_V\,,
\end{multline}
in which $d \mathcal{PS}_V$ is the phase space measure of the $V$ decay, 
and the amplitude for $B \to (\rho \to 2\pi) l \bar\nu$ or $B \to (\omega \to 3\pi) l \bar\nu$ decomposes in the narrow width approximation as
\begin{equation}
	\label{eq:prefamplfull}
	\mathcal{A}_{s_l s_\nu} = \sum_{\lambda_V} \frac{A_{\lambda_V s_l s_\nu} A^V_{\lambda_V}}{\sqrt{2 m_V\Gamma_V}} \,.
\end{equation}

The $\rho \to \pi\pi$ strong decay is generated via the chiral interaction $g_\rho \rho_\mu [\pi (\partial^\mu \pi) - (\partial^\mu \pi) \pi]$.
In the $\rho^0 \to \pi^+\pi^-$ ($\rho^- \to \pi^-\pi^0$) decay, we denote the momentum of the $\pi^+$ ($\pi^-$) in the $\rho^0$ ($\rho^-$) rest frame by $\bm{p}^*_{+}$ ($\bm{p}^*_{-}$), 
with magnitude $|p^*_\pi|$.
In our phase conventions, the $\rho \to \pi\pi$ amplitude is then
\begin{subequations}
\label{eqn:rhoampls}
\begin{align}
	A^\rho_\pm & = -g_\rho\sqrt{2}|p^*_\pi| e^{\pm i(\phi_\pi-\phtau)} \sin \theta_\pi \\
	A^\rho_0 & = 2g_\rho|p^*_\pi| \cos \theta_\pi\,, 
\end{align}
\end{subequations}
in which the helicity angles, $\theta_\pi$ and $\phi_\pi$, define the orientation of $\bm{p}^*_{+}$ ($\bm{p}^*_{-}$)
with respect to $+\bm{p}_{B}$ in the $\rho^0$ ($\rho^-$) rest frame. See Fig.~\ref{fig:hadef}.
Note that the physical twist angle $\phi_\pi-\phtau$ appears.  

Combining Eq.~\eqref{eq:prefamplfull} with Eqs.~\eqref{eqn:buampls} and~\eqref{eqn:rhoampls} yields the full amplitude expressions, 
from which square matrix elements follow immediately.
The phase space measure of the $\rho \to\pi\pi$ decay is trivially 
\begin{equation}
	d \mathcal{PS}_\rho = \frac{|p_\pi^*|}{16 \pi^2 m_\rho} d \Omega_\pi\,, 
\end{equation}	
over which integration of the square amplitudes is straightforward.
(One finds $\Gamma[\rho\to\pi\pi] = g_\rho^2 |p_\pi^*|^3/(2 \pi m_\rho^2)$.)
From Eqs.~\eqref{eqn:rhoampls} one may also immediately derive the differential decay in the cascade $B \to(\rho \to \pi\pi)l\nu$,
\begin{equation}
	\label{eqn:FLrhodiff}
	\frac{1}{\Gamma} \frac{d \Gamma}{d \cos\theta_\pi} = \frac{3}{2}\bigg[\big[1 - F_{L}(\rho)\big] \frac{\sin^2\theta_\pi}{2} + F_{L}(\rho) \cos^2\theta_\pi\bigg]
\end{equation}
in which $F_{L}(\rho)$ is the longitudinal polarization~\eqref{eqn:FLrho}.

The $\omega \to \pi^+ \pi^- \pi^0$ strong decay is generated via the interaction $g_\omega\varepsilon^{\mu\nu\rho\sigma}\omega_\mu \partial_\nu \pi \partial_\rho \pi \partial_\sigma \pi$.
For the $\omega \to \pi^+ \pi^- \pi^0$ decay, we denote the momenta of the $\pi^{\pm}$ in the $\omega$ rest frame by $\bm{p}^*_{\pm}$, with magnitude $|p^*_\pm|$, respectively.
In our phase conventions, the $\omega \to \pi^+ \pi^- \pi^0$ decay amplitude is then
\begin{subequations}
\label{eqn:omegaampls}
\begin{align}
	A^\omega_\pm  & =  \frac{g_\omega m_\omega |p^*_{+}| |p^*_{-}|}{\sqrt{2}} \bigg[e^{\pm i(\phi_{\pm} -\phtau)} \cos\theta_{\mp} \sin \theta_{\pm} \nn \\
		& \qquad \qquad  -e^{\pm i(\phi_{\mp} -\phtau)} \cos\theta_{\pm} \sin \theta_{\mp}\bigg] \\ 
	A^\omega_0  & = ig_\omega m_\omega |p^*_{+}| |p^*_{-}| \sin(\phi_{+} - \phi_{-}) \sin \theta_{+}\sin \theta_{-}\,.
\end{align}
\end{subequations}
Here the helicity angles, $\theta_{\pm}$ and $\phi_{\pm}$, define the orientation of $\bm{p}_{\pm}$
with respect to $+\bm{p}_{B}$ in the $\omega$ rest frame. See Fig.~\ref{fig:hadef}.
Note that two physical twist angles $\phi_{\pm}-\phtau$ appear.  

\begin{figure}[tb]
	\includegraphics[width = 0.55\linewidth]{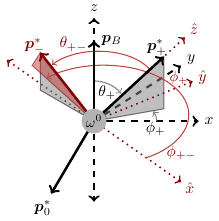}
	\caption{Definition of the relative helicity angles $\theta_{+-}$ and $\phi_{+-}$ with respect to the hatted coordinate system, shown in red, in the $\omega$ rest frame.
	The $\hat{z}$ axis aligns with $\bm{p}_+^*$; the $\hat{y}$ axis lies in the $x$--$y$ plane, at $-\phi_+$ from the $y$ axis. 
	The latter defines the orientation of the azimuthal angle $\phi_{+-}$ of $\bm{p}_-^*$ around $\bm{p}_+^*$. 
	The polar angle $\theta_{+-}$ is simply the angle between $\bm{p}_+^*$ and $\bm{p}_-^*$.}
	\label{fig:eulercoords}
\end{figure}

Combining Eq.~\eqref{eq:prefamplfull} with Eqs.~\eqref{eqn:buampls} and~\eqref{eqn:omegaampls} yields the full amplitude expressions.
However, for the $\omega \to 3\pi$ decay, the orientations of $\bm{p}^*_\pm$ cannot be chosen freely simultaneously, 
because $p_\omega - p_+ - p_- = p_0$ is constrained to be on the $\pi^0$ mass shell.
That is, one cannot simply integrate the square amplitude arising from Eq.~\eqref{eqn:omegaampls}
over $d\Omega_+d\Omega_-$, because the integration limits become non-trivial.

Natural coordinates for integration of the $\pi^+\pi^-\pi^0$ phase space may instead be constructed 
by defining relative polar coordinates $\theta_{+-}$ and $\phi_{+-}$ for, e.g., the $\pi^-$ with respect to the $\pi^+$,
in the usual spirit of a Dalitz-style analysis.
In particular, we choose coordinates as shown in Fig.~\ref{fig:eulercoords}, such that the $\hat{z}$ axis aligns with $\bm{p}_+^*$ 
and the $\hat{y}$ axis lies in the $x$--$y$ plane, at $-\phi_+$ from the $y$ axis.
The latter defines the azimuthal angle $\phi_{+-}$ of $\bm{p}^*_-$ around $\bm{p}^*_+$, while $\theta_{+-}$ is the angle between $\bm{p}^*_+$ and $\bm{p}^*_-$.
Because $d\Omega_+d\Omega_- = d\Omega_+d\Omega_{+-}$, the phase space measure becomes
\begin{equation}
	d \mathcal{PS}_\omega = \frac{dE^*_+ dE^*_-}{8(2\pi)^5} d\Omega_+ d\phi_{+-}\,,
\end{equation}
with the mass-shell constraint 
\begin{multline}
	\label{eqn:massshell}
	2|p^*_+||p^*_-|\cos\theta_{+-} = m_\omega^2 + 2m_+^2 - m_0^2 \\ - 2m_\omega (E^*_+  + E^*_-) + 2 E^*_+ E^*_-\,,
\end{multline}
in which $E^*_{\pm}$ are the energies of the $\pi^\pm$ in the $\omega$ rest-frame, 
and $m_+$ and $m_0$ are the $\pi^\pm$ and $\pi^0$ masses, respectively.
The integration domain of $dE^*_+ dE^*_-$ is non-trivial. 
However, defining $s = (p_+ + p_-)^2$ and $\mathcal{E}^\pm = E^*_{+} \pm E^*_{-}$, the measure can be further rewritten
\begin{equation}
	d \mathcal{PS}_\omega = \frac{ds d\mathcal{E}^-}{32 m_\omega(2\pi)^5} d\Omega_+ d\phi_{+-}\,,
\end{equation}
in which the ordered integration domain $4m_+^2 \le s \le (m_\omega - m_0)^2$ and $-\mathcal{E}^-_{\rm max} (s) \le \mathcal{E}_- \le \mathcal{E}^-_{\rm max}(s)$ with
\begin{align}
	\mathcal{E}^+(s) & = \frac{m_\omega^2 - m_0^2 + s}{2m_\omega}\,, \nn \\
	\mathcal{E}^-_{\rm max}(s) & = \Big[\big(s - 4 m_+^2\big)\Big(\mathcal{E}^+(s)^2/s - 1\Big)\Big]^{1/2}\,.
\end{align}

In these polar coordinates, the $\omega \to \pi^+\pi^-\pi^0$ helicity amplitudes become
\begin{subequations}
\label{eqn:omegaamplsEu}
\begin{align}
	A^\omega_\pm  & =  \mp\frac{g_\omega m_\omega |p^*_{+}| |p^*_{-}|}{\sqrt{2}} e^{\pm i(\phi_+ - \phtau)}  \nn \\
	& \quad \times \bigg[ \cos\phi_{+-} \pm i  \cos \theta_+ \sin \phi_{+-}\bigg]\sin \theta_{+-} \,,\\ 
	A^\omega_0  & =  -i g_\omega m_\omega |p^*_{+}| |p^*_{-}| \sin(\phi_{+-}) \sin \theta_{+}\sin \theta_{+-}\,.
\end{align}
\end{subequations}
Noting further from Eq.~\eqref{eqn:massshell}
\begin{equation}
	|p^*_{+}|^2 |p^*_{-}|^2\sin^2\theta_{+-} = \frac{s}{4}\Big[\mathcal{E}^-_{\rm max} (s)^2 - (\mathcal{E}^-)^2\Big]\,,
\end{equation}
integration of the square of the amplitudes over $d \mathcal{PS}_\omega$ is now straightforward.
(One finds $\Gamma[\omega \to 3\pi]  \simeq 1.94 \times 8/(6 \pi^3) g_\omega^2 m_\omega^7$.)
One may also immediately derive the differential decay in the cascade $B \to(\omega \to \pi\pi\pi)l\bar\nu$,
\begin{equation}
	\label{eqn:FLomegadiff}
	\frac{1}{\Gamma} \frac{d \Gamma}{d \cos\theta_+} = \frac{3}{8}\bigg[\big[1 - F_{L}(\omega)\big](1 + \cos^2\theta_+) + 2F_{L}(\omega) \sin^2\theta_+\bigg]
\end{equation}
in which $F_{L}(\omega)$ is the longitudinal polarization of the $\omega$.

\section{Correlations}
We give the post-fit correlation matrices for the spectrum average discussed in Sec.~\ref{sec:average} in Tables~\ref{tab:legacy-spectrum-correlation-rho} and~\ref{tab:legacy-spectrum-correlation-omega}. The post-fit correlation matrices for the form factor fits discussed in Sec.~\ref{sec:theo_fit} are provided in Tables~\ref{tab:fit-correlation-rho} and~\ref{tab:fit-correlation-omega}.

\onecolumngrid

\begin{table*}[tb]
	\caption{Correlation matrix of the averaged $\BtoRhoLNu$ spectrum.}
	\label{tab:legacy-spectrum-correlation-rho}
\scalebox{0.85}{
\begin{tabular}{crrrrrrrrrrr}
	\hline
	\hline	
	&   [0, 2] &   [2, 4] &   [4, 6] &   [6, 8] &   [8, 10] &   [10, 12] &   [12, 14] &   [14, 16] &   [16, 18] &   [18, 20] &   [20, 22] \\
	\hline
	{[0, 2]}   &     1.00 &    -0.30 &     0.03 &     0.01 &      0.09 &       0.09 &       0.09 &       0.09 &       0.08 &       0.08 &       0.02 \\
	{[2, 4]}   &    -0.30 &     1.00 &    -0.03 &     0.09 &      0.11 &       0.12 &       0.12 &       0.12 &       0.11 &       0.10 &       0.02 \\
	{[4, 6]}   &     0.03 &    -0.03 &     1.00 &    -0.18 &      0.13 &       0.13 &       0.15 &       0.14 &       0.13 &       0.12 &       0.03 \\
	{[6, 8]}   &     0.01 &     0.09 &    -0.18 &     1.00 &      0.06 &       0.18 &       0.18 &       0.18 &       0.16 &       0.14 &       0.04 \\
	{[8, 10]}  &     0.09 &     0.11 &     0.13 &     0.06 &      1.00 &      -0.21 &       0.05 &       0.04 &       0.12 &       0.10 &       0.03 \\
	{[10, 12]} &     0.09 &     0.12 &     0.13 &     0.18 &     -0.21 &       1.00 &      -0.00 &       0.07 &       0.15 &       0.13 &       0.04 \\
	{[12, 14]} &     0.09 &     0.12 &     0.15 &     0.18 &      0.05 &      -0.00 &       1.00 &      -0.16 &       0.14 &       0.12 &       0.04 \\
	{[14, 16]} &     0.09 &     0.12 &     0.14 &     0.18 &      0.04 &       0.07 &      -0.16 &       1.00 &       0.10 &       0.14 &       0.05 \\
	{[16, 18]} &     0.08 &     0.11 &     0.13 &     0.16 &      0.12 &       0.15 &       0.14 &       0.10 &       1.00 &      -0.27 &      -0.11 \\
	{[18, 20]} &     0.08 &     0.10 &     0.12 &     0.14 &      0.10 &       0.13 &       0.12 &       0.14 &      -0.27 &       1.00 &      -0.13 \\
	{[20, 22]} &     0.02 &     0.02 &     0.03 &     0.04 &      0.03 &       0.04 &       0.04 &       0.05 &      -0.11 &      -0.13 &       1.00 \\
	\hline
	\hline	
\end{tabular}
}
\end{table*}

\begin{table*}[tb]
	\caption{Correlation matrix of the averaged $\BtoOmegaLNu$ spectrum.}
	\label{tab:legacy-spectrum-correlation-omega}
\scalebox{0.85}{
\begin{tabular}{crrrrrrr}
	\hline
	\hline	
	&   [0, 4] &   [4, 8] &   [8, 10] &   [10, 12] &   [12, 21] &   $\theta_2$ &   $\theta_5$ \\
	\hline
	{[0, 4]}     &     1.00 &    -0.15 &      0.08 &       0.04 &       0.06 &        -0.01 &         0.00 \\
	{[4, 8]}     &    -0.15 &     1.00 &      0.09 &       0.09 &       0.15 &        -0.01 &        -0.00 \\
	{[8, 10]}    &     0.08 &     0.09 &      1.00 &      -0.01 &       0.12 &        -0.00 &        -0.00 \\
	{[10, 12]}   &     0.04 &     0.09 &     -0.01 &       1.00 &       0.15 &         0.00 &        -0.00 \\
	{[12, 21]}   &     0.06 &     0.15 &      0.12 &       0.15 &       1.00 &        -0.00 &        -0.00 \\
	$\theta_2$ &    -0.01 &    -0.01 &     -0.00 &       0.00 &      -0.00 &         1.00 &         0.00 \\
	$\theta_5$ &     0.00 &    -0.00 &     -0.00 &      -0.00 &      -0.00 &         0.00 &         1.00 \\
	\hline
	\hline	
\end{tabular}
}
\end{table*}

\begin{table*}[tb]
	\caption{Correlation matrix for $|\Vub|$ and the BSZ parameters to the averaged $\BtoRhoLNu$ spectrum and the LCSR data.}
	\label{tab:fit-correlation-rho}
	\resizebox{0.95\textwidth}{!}{%
\begin{tabular}{crrrrrrrrrrrrrrrrrrrr}
	\hline
	\hline
	&   $|V_\mathrm{ub}|$ &   $\alpha_1^{A_0}$ &   $\alpha_2^{A_0}$ &   $\alpha_0^{A_1}$ &   $\alpha_1^{A_1}$ &   $\alpha_2^{A_1}$ &   $\alpha_0^{A_{12}}$ &   $\alpha_1^{A_{12}}$ &   $\alpha_2^{A_{12}}$ &   $\alpha_0^{V}$ &   $\alpha_1^{V}$ &   $\alpha_2^{V}$ &   $\alpha_0^{T_1}$ &   $\alpha_1^{T_1}$ &   $\alpha_2^{T_1}$ &   $\alpha_1^{T_2}$ &   $\alpha_2^{T_2}$ &   $\alpha_0^{T_{23}}$ &   $\alpha_1^{T_{23}}$ &   $\alpha_2^{T_{23}}$ \\
	\hline
	$|V_\mathrm{ub}|$   &                1.00 &              -0.05 &              -0.02 &              -0.54 &               0.07 &               0.05 &                 -0.75 &                 -0.08 &                  0.04 &            -0.53 &             0.09 &            -0.02 &              -0.50 &               0.10 &              -0.03 &               0.08 &               0.07 &                 -0.55 &                 -0.11 &                 -0.01 \\
	$\alpha_1^{A_0}$    &               -0.05 &               1.00 &              -0.15 &               0.06 &               0.14 &               0.20 &                  0.30 &                  0.86 &                  0.81 &            -0.03 &             0.22 &             0.16 &              -0.04 &               0.24 &               0.12 &               0.15 &               0.27 &                  0.26 &                  0.86 &                  0.73 \\
	$\alpha_2^{A_0}$    &               -0.02 &              -0.15 &               1.00 &               0.02 &               0.14 &               0.28 &                 -0.07 &                 -0.23 &                 -0.19 &            -0.02 &             0.06 &             0.57 &              -0.17 &               0.06 &               0.56 &               0.11 &               0.44 &                 -0.03 &                  0.07 &                  0.40 \\
	$\alpha_0^{A_1}$    &               -0.54 &               0.06 &               0.02 &               1.00 &               0.56 &               0.46 &                  0.29 &                  0.02 &                 -0.15 &             0.90 &             0.54 &            -0.31 &               0.88 &               0.54 &              -0.33 &               0.55 &               0.38 &                  0.24 &                  0.20 &                  0.09 \\
	$\alpha_1^{A_1}$    &                0.07 &               0.14 &               0.14 &               0.56 &               1.00 &               0.87 &                 -0.13 &                  0.08 &                 -0.04 &             0.48 &             0.95 &            -0.24 &               0.45 &               0.95 &              -0.28 &               0.98 &               0.79 &                 -0.06 &                  0.23 &                  0.18 \\
	$\alpha_2^{A_1}$    &                0.05 &               0.20 &               0.28 &               0.46 &               0.87 &               1.00 &                 -0.12 &                  0.12 &                  0.08 &             0.37 &             0.88 &             0.02 &               0.31 &               0.88 &              -0.03 &               0.87 &               0.94 &                 -0.04 &                  0.28 &                  0.32 \\
	$\alpha_0^{A_{12}}$ &               -0.75 &               0.30 &              -0.07 &               0.29 &              -0.13 &              -0.12 &                  1.00 &                  0.44 &                  0.26 &             0.28 &            -0.15 &             0.06 &               0.26 &              -0.15 &               0.07 &              -0.13 &              -0.14 &                  0.69 &                  0.32 &                  0.12 \\
	$\alpha_1^{A_{12}}$ &               -0.08 &               0.86 &              -0.23 &               0.02 &               0.08 &               0.12 &                  0.44 &                  1.00 &                  0.89 &            -0.05 &             0.19 &            -0.06 &              -0.02 &               0.21 &              -0.09 &               0.09 &               0.11 &                  0.25 &                  0.80 &                  0.59 \\
	$\alpha_2^{A_{12}}$ &                0.04 &               0.81 &              -0.19 &              -0.15 &              -0.04 &               0.08 &                  0.26 &                  0.89 &                  1.00 &            -0.23 &             0.09 &             0.04 &              -0.19 &               0.11 &              -0.00 &              -0.01 &               0.10 &                  0.08 &                  0.60 &                  0.57 \\
	$\alpha_0^{V}$      &               -0.53 &              -0.03 &              -0.02 &               0.90 &               0.48 &               0.37 &                  0.28 &                 -0.05 &                 -0.23 &             1.00 &             0.54 &            -0.37 &               0.90 &               0.48 &              -0.34 &               0.50 &               0.31 &                  0.23 &                  0.11 &                 -0.00 \\
	$\alpha_1^{V}$      &                0.09 &               0.22 &               0.06 &               0.54 &               0.95 &               0.88 &                 -0.15 &                  0.19 &                  0.09 &             0.54 &             1.00 &            -0.33 &               0.48 &               0.97 &              -0.36 &               0.96 &               0.80 &                 -0.11 &                  0.29 &                  0.22 \\
	$\alpha_2^{V}$      &               -0.02 &               0.16 &               0.57 &              -0.31 &              -0.24 &               0.02 &                  0.06 &                 -0.06 &                  0.04 &            -0.37 &            -0.33 &             1.00 &              -0.47 &              -0.32 &               0.96 &              -0.29 &               0.22 &                  0.15 &                  0.24 &                  0.52 \\
	$\alpha_0^{T_1}$    &               -0.50 &              -0.04 &              -0.17 &               0.88 &               0.45 &               0.31 &                  0.26 &                 -0.02 &                 -0.19 &             0.90 &             0.48 &            -0.47 &               1.00 &               0.51 &              -0.53 &               0.50 &               0.23 &                  0.20 &                  0.07 &                 -0.10 \\
	$\alpha_1^{T_1}$    &                0.10 &               0.24 &               0.06 &               0.54 &               0.95 &               0.88 &                 -0.15 &                  0.21 &                  0.11 &             0.48 &             0.97 &            -0.32 &               0.51 &               1.00 &              -0.39 &               0.98 &               0.82 &                 -0.10 &                  0.31 &                  0.24 \\
	$\alpha_2^{T_1}$    &               -0.03 &               0.12 &               0.56 &              -0.33 &              -0.28 &              -0.03 &                  0.07 &                 -0.09 &                 -0.00 &            -0.34 &            -0.36 &             0.96 &              -0.53 &              -0.39 &               1.00 &              -0.34 &               0.15 &                  0.17 &                  0.20 &                  0.47 \\
	$\alpha_1^{T_2}$    &                0.08 &               0.15 &               0.11 &               0.55 &               0.98 &               0.87 &                 -0.13 &                  0.09 &                 -0.01 &             0.50 &             0.96 &            -0.29 &               0.50 &               0.98 &              -0.34 &               1.00 &               0.81 &                 -0.07 &                  0.23 &                  0.18 \\
	$\alpha_2^{T_2}$    &                0.07 &               0.27 &               0.44 &               0.38 &               0.79 &               0.94 &                 -0.14 &                  0.11 &                  0.10 &             0.31 &             0.80 &             0.22 &               0.23 &               0.82 &               0.15 &               0.81 &               1.00 &                 -0.04 &                  0.36 &                  0.50 \\
	$\alpha_0^{T_{23}}$ &               -0.55 &               0.26 &              -0.03 &               0.24 &              -0.06 &              -0.04 &                  0.69 &                  0.25 &                  0.08 &             0.23 &            -0.11 &             0.15 &               0.20 &              -0.10 &               0.17 &              -0.07 &              -0.04 &                  1.00 &                  0.32 &                  0.11 \\
	$\alpha_1^{T_{23}}$ &               -0.11 &               0.86 &               0.07 &               0.20 &               0.23 &               0.28 &                  0.32 &                  0.80 &                  0.60 &             0.11 &             0.29 &             0.24 &               0.07 &               0.31 &               0.20 &               0.23 &               0.36 &                  0.32 &                  1.00 &                  0.86 \\
	$\alpha_2^{T_{23}}$ &               -0.01 &               0.73 &               0.40 &               0.09 &               0.18 &               0.32 &                  0.12 &                  0.59 &                  0.57 &            -0.00 &             0.22 &             0.52 &              -0.10 &               0.24 &               0.47 &               0.18 &               0.50 &                  0.11 &                  0.86 &                  1.00 \\
	\hline
	\hline
\end{tabular}}
\end{table*}

\begin{table*}[tb]
	\caption{Correlation matrix for $|\Vub|$ and the BSZ parameters to the averaged $\BtoOmegaLNu$ spectrum and the LCSR data.}
	\label{tab:fit-correlation-omega}
	\resizebox{0.95\textwidth}{!}{%
\begin{tabular}{crrrrrrrrrrrrrrrrrrrr}
	\hline
	\hline
	&   $|V_\mathrm{ub}|$ &   $\alpha_1^{A_0}$ &   $\alpha_2^{A_0}$ &   $\alpha_0^{A_1}$ &   $\alpha_1^{A_1}$ &   $\alpha_2^{A_1}$ &   $\alpha_0^{A_{12}}$ &   $\alpha_1^{A_{12}}$ &   $\alpha_2^{A_{12}}$ &   $\alpha_0^{V}$ &   $\alpha_1^{V}$ &   $\alpha_2^{V}$ &   $\alpha_0^{T_1}$ &   $\alpha_1^{T_1}$ &   $\alpha_2^{T_1}$ &   $\alpha_1^{T_2}$ &   $\alpha_2^{T_2}$ &   $\alpha_0^{T_{23}}$ &   $\alpha_1^{T_{23}}$ &   $\alpha_2^{T_{23}}$ \\
	\hline
	$|V_\mathrm{ub}|$   &                1.00 &              -0.22 &               0.08 &              -0.48 &               0.04 &               0.04 &                 -0.80 &                 -0.28 &                 -0.20 &            -0.46 &             0.06 &            -0.04 &              -0.43 &               0.06 &              -0.05 &               0.05 &               0.05 &                 -0.61 &                 -0.24 &                 -0.19 \\
	$\alpha_1^{A_0}$    &               -0.22 &               1.00 &              -0.48 &               0.12 &               0.04 &               0.03 &                  0.47 &                  0.93 &                  0.85 &             0.05 &             0.11 &             0.08 &               0.08 &               0.14 &               0.03 &               0.04 &               0.08 &                  0.49 &                  0.92 &                  0.81 \\
	$\alpha_2^{A_0}$    &                0.08 &              -0.48 &               1.00 &               0.03 &               0.05 &               0.10 &                 -0.27 &                 -0.52 &                 -0.39 &             0.07 &            -0.01 &             0.18 &              -0.02 &              -0.02 &               0.21 &               0.05 &               0.16 &                 -0.30 &                 -0.40 &                 -0.10 \\
	$\alpha_0^{A_1}$    &               -0.48 &               0.12 &               0.03 &               1.00 &               0.61 &               0.46 &                  0.24 &                  0.05 &                 -0.10 &             0.94 &             0.60 &            -0.46 &               0.93 &               0.61 &              -0.48 &               0.61 &               0.45 &                  0.26 &                  0.21 &                  0.13 \\
	$\alpha_1^{A_1}$    &                0.04 &               0.04 &               0.05 &               0.61 &               1.00 &               0.84 &                 -0.08 &                 -0.02 &                 -0.24 &             0.59 &             0.97 &            -0.51 &               0.58 &               0.97 &              -0.55 &               0.99 &               0.81 &                 -0.02 &                  0.12 &                 -0.05 \\
	$\alpha_2^{A_1}$    &                0.04 &               0.03 &               0.10 &               0.46 &               0.84 &               1.00 &                 -0.07 &                 -0.00 &                 -0.13 &             0.41 &             0.87 &            -0.16 &               0.40 &               0.86 &              -0.22 &               0.84 &               0.95 &                 -0.02 &                  0.09 &                  0.01 \\
	$\alpha_0^{A_{12}}$ &               -0.80 &               0.47 &              -0.27 &               0.24 &              -0.08 &              -0.07 &                  1.00 &                  0.59 &                  0.46 &             0.21 &            -0.10 &             0.12 &               0.20 &              -0.09 &               0.13 &              -0.09 &              -0.08 &                  0.76 &                  0.47 &                  0.32 \\
	$\alpha_1^{A_{12}}$ &               -0.28 &               0.93 &              -0.52 &               0.05 &              -0.02 &              -0.00 &                  0.59 &                  1.00 &                  0.89 &            -0.01 &             0.06 &             0.05 &               0.02 &               0.08 &               0.01 &              -0.02 &               0.01 &                  0.51 &                  0.89 &                  0.75 \\
	$\alpha_2^{A_{12}}$ &               -0.20 &               0.85 &              -0.39 &              -0.10 &              -0.24 &              -0.13 &                  0.46 &                  0.89 &                  1.00 &            -0.16 &            -0.14 &             0.19 &              -0.13 &              -0.12 &               0.16 &              -0.22 &              -0.10 &                  0.36 &                  0.70 &                  0.73 \\
	$\alpha_0^{V}$      &               -0.46 &               0.05 &               0.07 &               0.94 &               0.59 &               0.41 &                  0.21 &                 -0.01 &                 -0.16 &             1.00 &             0.61 &            -0.52 &               0.93 &               0.59 &              -0.50 &               0.60 &               0.41 &                  0.21 &                  0.14 &                  0.07 \\
	$\alpha_1^{V}$      &                0.06 &               0.11 &              -0.01 &               0.60 &               0.97 &               0.87 &                 -0.10 &                  0.06 &                 -0.14 &             0.61 &             1.00 &            -0.52 &               0.59 &               0.99 &              -0.56 &               0.98 &               0.85 &                 -0.03 &                  0.18 &                  0.01 \\
	$\alpha_2^{V}$      &               -0.04 &               0.08 &               0.18 &              -0.46 &              -0.51 &              -0.16 &                  0.12 &                  0.05 &                  0.19 &            -0.52 &            -0.52 &             1.00 &              -0.54 &              -0.51 &               0.95 &              -0.53 &              -0.07 &                  0.13 &                  0.08 &                  0.30 \\
	$\alpha_0^{T_1}$    &               -0.43 &               0.08 &              -0.02 &               0.93 &               0.58 &               0.40 &                  0.20 &                  0.02 &                 -0.13 &             0.93 &             0.59 &            -0.54 &               1.00 &               0.62 &              -0.60 &               0.61 &               0.39 &                  0.21 &                  0.15 &                  0.04 \\
	$\alpha_1^{T_1}$    &                0.06 &               0.14 &              -0.02 &               0.61 &               0.97 &               0.86 &                 -0.09 &                  0.08 &                 -0.12 &             0.59 &             0.99 &            -0.51 &               0.62 &               1.00 &              -0.58 &               0.99 &               0.85 &                 -0.02 &                  0.21 &                  0.03 \\
	$\alpha_2^{T_1}$    &               -0.05 &               0.03 &               0.21 &              -0.48 &              -0.55 &              -0.22 &                  0.13 &                  0.01 &                  0.16 &            -0.50 &            -0.56 &             0.95 &              -0.60 &              -0.58 &               1.00 &              -0.59 &              -0.13 &                  0.11 &                  0.03 &                  0.26 \\
	$\alpha_1^{T_2}$    &                0.05 &               0.04 &               0.05 &               0.61 &               0.99 &               0.84 &                 -0.09 &                 -0.02 &                 -0.22 &             0.60 &             0.98 &            -0.53 &               0.61 &               0.99 &              -0.59 &               1.00 &               0.82 &                 -0.03 &                  0.12 &                 -0.05 \\
	$\alpha_2^{T_2}$    &                0.05 &               0.08 &               0.16 &               0.45 &               0.81 &               0.95 &                 -0.08 &                  0.01 &                 -0.10 &             0.41 &             0.85 &            -0.07 &               0.39 &               0.85 &              -0.13 &               0.82 &               1.00 &                 -0.02 &                  0.16 &                  0.13 \\
	$\alpha_0^{T_{23}}$ &               -0.61 &               0.49 &              -0.30 &               0.26 &              -0.02 &              -0.02 &                  0.76 &                  0.51 &                  0.36 &             0.21 &            -0.03 &             0.13 &               0.21 &              -0.02 &               0.11 &              -0.03 &              -0.02 &                  1.00 &                  0.53 &                  0.32 \\
	$\alpha_1^{T_{23}}$ &               -0.24 &               0.92 &              -0.40 &               0.21 &               0.12 &               0.09 &                  0.47 &                  0.89 &                  0.70 &             0.14 &             0.18 &             0.08 &               0.15 &               0.21 &               0.03 &               0.12 &               0.16 &                  0.53 &                  1.00 &                  0.86 \\
	$\alpha_2^{T_{23}}$ &               -0.19 &               0.81 &              -0.10 &               0.13 &              -0.05 &               0.01 &                  0.32 &                  0.75 &                  0.73 &             0.07 &             0.01 &             0.30 &               0.04 &               0.03 &               0.26 &              -0.05 &               0.13 &                  0.32 &                  0.86 &                  1.00 \\
	\hline
	\hline
\end{tabular}}
\end{table*}

\FloatBarrier
\twocolumngrid

\bibliographystyle{apsrev4-1}
\bibliography{paper}

\begin{thebibliography}{38}%
\makeatletter
\providecommand \@ifxundefined [1]{%
 \@ifx{#1\undefined}
}%
\providecommand \@ifnum [1]{%
 \ifnum #1\expandafter \@firstoftwo
 \else \expandafter \@secondoftwo
 \fi
}%
\providecommand \@ifx [1]{%
 \ifx #1\expandafter \@firstoftwo
 \else \expandafter \@secondoftwo
 \fi
}%
\providecommand \natexlab [1]{#1}%
\providecommand \enquote  [1]{``#1''}%
\providecommand \bibnamefont  [1]{#1}%
\providecommand \bibfnamefont [1]{#1}%
\providecommand \citenamefont [1]{#1}%
\providecommand \href@noop [0]{\@secondoftwo}%
\providecommand \href [0]{\begingroup \@sanitize@url \@href}%
\providecommand \@href[1]{\@@startlink{#1}\@@href}%
\providecommand \@@href[1]{\endgroup#1\@@endlink}%
\providecommand \@sanitize@url [0]{\catcode `\\12\catcode `\$12\catcode
  `\&12\catcode `\#12\catcode `\^12\catcode `\_12\catcode `\%12\relax}%
\providecommand \@@startlink[1]{}%
\providecommand \@@endlink[0]{}%
\providecommand \url  [0]{\begingroup\@sanitize@url \@url }%
\providecommand \@url [1]{\endgroup\@href {#1}{\urlprefix }}%
\providecommand \urlprefix  [0]{URL }%
\providecommand \Eprint [0]{\href }%
\providecommand \doibase [0]{http://dx.doi.org/}%
\providecommand \selectlanguage [0]{\@gobble}%
\providecommand \bibinfo  [0]{\@secondoftwo}%
\providecommand \bibfield  [0]{\@secondoftwo}%
\providecommand \translation [1]{[#1]}%
\providecommand \BibitemOpen [0]{}%
\providecommand \bibitemStop [0]{}%
\providecommand \bibitemNoStop [0]{.\EOS\space}%
\providecommand \EOS [0]{\spacefactor3000\relax}%
\providecommand \BibitemShut  [1]{\csname bibitem#1\endcsname}%
\let\auto@bib@innerbib\@empty
\bibitem [{\citenamefont {Amhis}\ \emph {et~al.}(2019)\citenamefont {Amhis}
  \emph {et~al.}}]{Amhis:2019ckw}%
  \BibitemOpen
  \bibfield  {author} {\bibinfo {author} {\bibfnamefont {Y.~S.}\ \bibnamefont
  {Amhis}} \emph {et~al.} (\bibinfo {collaboration} {HFLAV}),\ }\href@noop {}
  {\  (\bibinfo {year} {2019})},\ \Eprint {http://arxiv.org/abs/1909.12524}
  {arXiv:1909.12524 [hep-ex]} \BibitemShut {NoStop}%
\bibitem [{\citenamefont {Bernlochner}\ \emph {et~al.}(2021)\citenamefont
  {Bernlochner}, \citenamefont {Sevilla}, \citenamefont {Robinson},\ and\
  \citenamefont {Wormser}}]{Bernlochner:2021vlv}%
  \BibitemOpen
  \bibfield  {author} {\bibinfo {author} {\bibfnamefont {F.~U.}\ \bibnamefont
  {Bernlochner}}, \bibinfo {author} {\bibfnamefont {M.~F.}\ \bibnamefont
  {Sevilla}}, \bibinfo {author} {\bibfnamefont {D.~J.}\ \bibnamefont
  {Robinson}}, \ and\ \bibinfo {author} {\bibfnamefont {G.}~\bibnamefont
  {Wormser}},\ }\href@noop {} {\  (\bibinfo {year} {2021})},\ \Eprint
  {http://arxiv.org/abs/2101.08326} {arXiv:2101.08326 [hep-ex]} \BibitemShut
  {NoStop}%
\bibitem [{\citenamefont {Bernlochner}(2015)}]{Bernlochner:2015mya}%
  \BibitemOpen
  \bibfield  {author} {\bibinfo {author} {\bibfnamefont {F.~U.}\ \bibnamefont
  {Bernlochner}},\ }\href {\doibase 10.1103/PhysRevD.92.115019} {\bibfield
  {journal} {\bibinfo  {journal} {Phys. Rev. D}\ }\textbf {\bibinfo {volume}
  {92}},\ \bibinfo {pages} {115019} (\bibinfo {year} {2015})},\ \Eprint
  {http://arxiv.org/abs/1509.06938} {arXiv:1509.06938 [hep-ph]} \BibitemShut
  {NoStop}%
\bibitem [{\citenamefont {Be\v{c}irevi\'c}\ \emph {et~al.}(2020)\citenamefont
  {Be\v{c}irevi\'c}, \citenamefont {Jaffredo}, \citenamefont {Pe\~nuelas},\
  and\ \citenamefont {Sumensari}}]{Becirevic:2020rzi}%
  \BibitemOpen
  \bibfield  {author} {\bibinfo {author} {\bibfnamefont {D.}~\bibnamefont
  {Be\v{c}irevi\'c}}, \bibinfo {author} {\bibfnamefont {F.}~\bibnamefont
  {Jaffredo}}, \bibinfo {author} {\bibfnamefont {A.}~\bibnamefont
  {Pe\~nuelas}}, \ and\ \bibinfo {author} {\bibfnamefont {O.}~\bibnamefont
  {Sumensari}},\ }\href@noop {} {\  (\bibinfo {year} {2020})},\ \Eprint
  {http://arxiv.org/abs/2012.09872} {arXiv:2012.09872 [hep-ph]} \BibitemShut
  {NoStop}%
\bibitem [{\citenamefont {Leljak}\ \emph {et~al.}(2021)\citenamefont {Leljak},
  \citenamefont {van Dyk},\ and\ \citenamefont {Meli\'c}}]{Leljak:2021vte}%
  \BibitemOpen
  \bibfield  {author} {\bibinfo {author} {\bibfnamefont {D.}~\bibnamefont
  {Leljak}}, \bibinfo {author} {\bibfnamefont {D.}~\bibnamefont {van Dyk}}, \
  and\ \bibinfo {author} {\bibfnamefont {B.}~\bibnamefont {Meli\'c}},\
  }\href@noop {} {\  (\bibinfo {year} {2021})},\ \Eprint
  {http://arxiv.org/abs/2102.07233} {arXiv:2102.07233 [hep-ph]} \BibitemShut
  {NoStop}%
\bibitem [{\citenamefont {Hamer}\ \emph {et~al.}(2016)\citenamefont {Hamer}
  \emph {et~al.}}]{Hamer:2015jsa}%
  \BibitemOpen
  \bibfield  {author} {\bibinfo {author} {\bibfnamefont {P.}~\bibnamefont
  {Hamer}} \emph {et~al.} (\bibinfo {collaboration} {Belle}),\ }\href {\doibase
  10.1103/PhysRevD.93.032007} {\bibfield  {journal} {\bibinfo  {journal} {Phys.
  Rev. D}\ }\textbf {\bibinfo {volume} {93}},\ \bibinfo {pages} {032007}
  (\bibinfo {year} {2016})},\ \Eprint {http://arxiv.org/abs/1509.06521}
  {arXiv:1509.06521 [hep-ex]} \BibitemShut {NoStop}%
\bibitem [{\citenamefont {Sibidanov}\ \emph {et~al.}(2013)\citenamefont
  {Sibidanov} \emph {et~al.}}]{Sibidanov:2013rkk}%
  \BibitemOpen
  \bibfield  {author} {\bibinfo {author} {\bibfnamefont {A.}~\bibnamefont
  {Sibidanov}} \emph {et~al.} (\bibinfo {collaboration} {Belle
  Collaboration}),\ }\href {\doibase 10.1103/PhysRevD.88.032005} {\bibfield
  {journal} {\bibinfo  {journal} {Phys. Rev.}\ }\textbf {\bibinfo {volume} {D
  88}},\ \bibinfo {pages} {032005} (\bibinfo {year} {2013})},\ \Eprint
  {http://arxiv.org/abs/1306.2781} {arXiv:1306.2781 [hep-ex]} \BibitemShut
  {NoStop}%
\bibitem [{\citenamefont {del Amo~Sanchez}\ \emph {et~al.}(2011)\citenamefont
  {del Amo~Sanchez} \emph {et~al.}}]{delAmoSanchez:2010af}%
  \BibitemOpen
  \bibfield  {author} {\bibinfo {author} {\bibfnamefont {P.}~\bibnamefont {del
  Amo~Sanchez}} \emph {et~al.} (\bibinfo {collaboration} {BaBar
  Collaboration}),\ }\href {\doibase 10.1103/PhysRevD.83.032007} {\bibfield
  {journal} {\bibinfo  {journal} {Phys. Rev.}\ }\textbf {\bibinfo {volume} {D
  83}},\ \bibinfo {pages} {032007} (\bibinfo {year} {2011})},\ \Eprint
  {http://arxiv.org/abs/1005.3288} {arXiv:1005.3288 [hep-ex]} \BibitemShut
  {NoStop}%
\bibitem [{\citenamefont {Lees}\ \emph {et~al.}(2013)\citenamefont {Lees} \emph
  {et~al.}}]{Lees:2012mq}%
  \BibitemOpen
  \bibfield  {author} {\bibinfo {author} {\bibfnamefont {J.~P.}\ \bibnamefont
  {Lees}} \emph {et~al.} (\bibinfo {collaboration} {BaBar}),\ }\href {\doibase
  10.1103/PhysRevD.87.099904, 10.1103/PhysRevD.87.032004} {\bibfield  {journal}
  {\bibinfo  {journal} {Phys. Rev.}\ }\textbf {\bibinfo {volume} {D87}},\
  \bibinfo {pages} {032004} (\bibinfo {year} {2013})},\ \bibinfo {note}
  {[Erratum: Phys. Rev.D87,no.9,099904(2013)]},\ \Eprint
  {http://arxiv.org/abs/1205.6245} {arXiv:1205.6245 [hep-ex]} \BibitemShut
  {NoStop}%
\bibitem [{\citenamefont {Bharucha}\ \emph {et~al.}(2016)\citenamefont
  {Bharucha}, \citenamefont {Straub},\ and\ \citenamefont
  {Zwicky}}]{Straub:2015ica}%
  \BibitemOpen
  \bibfield  {author} {\bibinfo {author} {\bibfnamefont {A.}~\bibnamefont
  {Bharucha}}, \bibinfo {author} {\bibfnamefont {D.~M.}\ \bibnamefont
  {Straub}}, \ and\ \bibinfo {author} {\bibfnamefont {R.}~\bibnamefont
  {Zwicky}},\ }\href {\doibase 10.1007/JHEP08(2016)098} {\bibfield  {journal}
  {\bibinfo  {journal} {JHEP}\ }\textbf {\bibinfo {volume} {08}},\ \bibinfo
  {pages} {098} (\bibinfo {year} {2016})},\ \Eprint
  {http://arxiv.org/abs/1503.05534} {arXiv:1503.05534 [hep-ph]} \BibitemShut
  {NoStop}%
\bibitem [{\citenamefont {Cao}\ \emph {et~al.}(2021)\citenamefont {Cao} \emph
  {et~al.}}]{Cao:2021xqf}%
  \BibitemOpen
  \bibfield  {author} {\bibinfo {author} {\bibfnamefont {L.}~\bibnamefont
  {Cao}} \emph {et~al.} (\bibinfo {collaboration} {Belle}),\ }\href@noop {} {\
  (\bibinfo {year} {2021})},\ \Eprint {http://arxiv.org/abs/2102.00020}
  {arXiv:2102.00020 [hep-ex]} \BibitemShut {NoStop}%
\bibitem [{\citenamefont {Bernlochner}\ \emph
  {et~al.}(2020{\natexlab{a}})\citenamefont {Bernlochner}, \citenamefont
  {Duell}, \citenamefont {Ligeti}, \citenamefont {Papucci},\ and\ \citenamefont
  {Robinson}}]{Bernlochner:2020tfi}%
  \BibitemOpen
  \bibfield  {author} {\bibinfo {author} {\bibfnamefont {F.~U.}\ \bibnamefont
  {Bernlochner}}, \bibinfo {author} {\bibfnamefont {S.}~\bibnamefont {Duell}},
  \bibinfo {author} {\bibfnamefont {Z.}~\bibnamefont {Ligeti}}, \bibinfo
  {author} {\bibfnamefont {M.}~\bibnamefont {Papucci}}, \ and\ \bibinfo
  {author} {\bibfnamefont {D.~J.}\ \bibnamefont {Robinson}},\ }\href {\doibase
  10.1140/epjc/s10052-020-8304-0} {\bibfield  {journal} {\bibinfo  {journal}
  {Eur. Phys. J. C}\ }\textbf {\bibinfo {volume} {80}},\ \bibinfo {pages} {883}
  (\bibinfo {year} {2020}{\natexlab{a}})},\ \Eprint
  {http://arxiv.org/abs/2002.00020} {arXiv:2002.00020 [hep-ph]} \BibitemShut
  {NoStop}%
\bibitem [{\citenamefont {Bernlochner}\ \emph
  {et~al.}(2020{\natexlab{b}})\citenamefont {Bernlochner}, \citenamefont
  {Duell}, \citenamefont {Ligeti}, \citenamefont {Papucci},\ and\ \citenamefont
  {Robinson}}]{bernlochner_florian_urs_2020_3993770}%
  \BibitemOpen
  \bibfield  {author} {\bibinfo {author} {\bibfnamefont {F.~U.}\ \bibnamefont
  {Bernlochner}}, \bibinfo {author} {\bibfnamefont {S.}~\bibnamefont {Duell}},
  \bibinfo {author} {\bibfnamefont {Z.}~\bibnamefont {Ligeti}}, \bibinfo
  {author} {\bibfnamefont {M.}~\bibnamefont {Papucci}}, \ and\ \bibinfo
  {author} {\bibfnamefont {D.~J.}\ \bibnamefont {Robinson}},\ }\href {\doibase
  10.5281/zenodo.3993770} {\enquote {\bibinfo {title} {{HAMMER - Helicity
  Amplitude Module for Matrix Element Reweighting}},}\ } (\bibinfo {year}
  {2020}{\natexlab{b}})\BibitemShut {NoStop}%
\bibitem [{\citenamefont {Kang}\ \emph {et~al.}(2014)\citenamefont {Kang},
  \citenamefont {Kubis}, \citenamefont {Hanhart},\ and\ \citenamefont
  {Mei\ss{}ner}}]{Kang:2013jaa}%
  \BibitemOpen
  \bibfield  {author} {\bibinfo {author} {\bibfnamefont {X.-W.}\ \bibnamefont
  {Kang}}, \bibinfo {author} {\bibfnamefont {B.}~\bibnamefont {Kubis}},
  \bibinfo {author} {\bibfnamefont {C.}~\bibnamefont {Hanhart}}, \ and\
  \bibinfo {author} {\bibfnamefont {U.-G.}\ \bibnamefont {Mei\ss{}ner}},\
  }\href {\doibase 10.1103/PhysRevD.89.053015} {\bibfield  {journal} {\bibinfo
  {journal} {Phys. Rev. D}\ }\textbf {\bibinfo {volume} {89}},\ \bibinfo
  {pages} {053015} (\bibinfo {year} {2014})},\ \Eprint
  {http://arxiv.org/abs/1312.1193} {arXiv:1312.1193 [hep-ph]} \BibitemShut
  {NoStop}%
\bibitem [{\citenamefont {Bele\~no}\ \emph {et~al.}(2020)\citenamefont
  {Bele\~no} \emph {et~al.}}]{Beleno:2020gzt}%
  \BibitemOpen
  \bibfield  {author} {\bibinfo {author} {\bibfnamefont {C.}~\bibnamefont
  {Bele\~no}} \emph {et~al.} (\bibinfo {collaboration} {Belle}),\ }\href@noop
  {} {\  (\bibinfo {year} {2020})},\ \Eprint {http://arxiv.org/abs/2005.07766}
  {arXiv:2005.07766 [hep-ex]} \BibitemShut {NoStop}%
\bibitem [{\citenamefont {Sirlin}(1982)}]{Sirlin:1981ie}%
  \BibitemOpen
  \bibfield  {author} {\bibinfo {author} {\bibfnamefont {A.}~\bibnamefont
  {Sirlin}},\ }\href {\doibase 10.1016/0550-3213(82)90303-0} {\bibfield
  {journal} {\bibinfo  {journal} {Nucl. Phys. B}\ }\textbf {\bibinfo {volume}
  {196}},\ \bibinfo {pages} {83} (\bibinfo {year} {1982})}\BibitemShut
  {NoStop}%
\bibitem [{\citenamefont {Tostado}\ and\ \citenamefont
  {L\'opez~Castro}(2016)}]{Tostado:2015tna}%
  \BibitemOpen
  \bibfield  {author} {\bibinfo {author} {\bibfnamefont {S.~L.}\ \bibnamefont
  {Tostado}}\ and\ \bibinfo {author} {\bibfnamefont {G.}~\bibnamefont
  {L\'opez~Castro}},\ }\href {\doibase 10.1140/epjc/s10052-016-4329-9}
  {\bibfield  {journal} {\bibinfo  {journal} {Eur. Phys. J. C}\ }\textbf
  {\bibinfo {volume} {76}},\ \bibinfo {pages} {495} (\bibinfo {year} {2016})},\
  \Eprint {http://arxiv.org/abs/1510.08020} {arXiv:1510.08020 [hep-ph]}
  \BibitemShut {NoStop}%
\bibitem [{\citenamefont {Boyd}\ \emph {et~al.}(1996)\citenamefont {Boyd},
  \citenamefont {Grinstein},\ and\ \citenamefont {Lebed}}]{Boyd:1995sq}%
  \BibitemOpen
  \bibfield  {author} {\bibinfo {author} {\bibfnamefont {C.}~\bibnamefont
  {Boyd}}, \bibinfo {author} {\bibfnamefont {B.}~\bibnamefont {Grinstein}}, \
  and\ \bibinfo {author} {\bibfnamefont {R.~F.}\ \bibnamefont {Lebed}},\ }\href
  {\doibase 10.1016/0550-3213(95)00653-2} {\bibfield  {journal} {\bibinfo
  {journal} {Nucl.\ Phys.\ B}\ }\textbf {\bibinfo {volume} {461}},\ \bibinfo
  {pages} {493} (\bibinfo {year} {1996})},\ \Eprint
  {http://arxiv.org/abs/hep-ph/9508211} {arXiv:hep-ph/9508211} \BibitemShut
  {NoStop}%
\bibitem [{\citenamefont {Boyd}\ \emph {et~al.}(1997)\citenamefont {Boyd},
  \citenamefont {Grinstein},\ and\ \citenamefont {Lebed}}]{Boyd:1997kz}%
  \BibitemOpen
  \bibfield  {author} {\bibinfo {author} {\bibfnamefont {C.}~\bibnamefont
  {Boyd}}, \bibinfo {author} {\bibfnamefont {B.}~\bibnamefont {Grinstein}}, \
  and\ \bibinfo {author} {\bibfnamefont {R.~F.}\ \bibnamefont {Lebed}},\ }\href
  {\doibase 10.1103/PhysRevD.56.6895} {\bibfield  {journal} {\bibinfo
  {journal} {Phys.\ Rev.\ D}\ }\textbf {\bibinfo {volume} {56}},\ \bibinfo
  {pages} {6895} (\bibinfo {year} {1997})},\ \Eprint
  {http://arxiv.org/abs/hep-ph/9705252} {arXiv:hep-ph/9705252} \BibitemShut
  {NoStop}%
\bibitem [{\citenamefont {Bourrely}\ \emph {et~al.}(2009)\citenamefont
  {Bourrely}, \citenamefont {Caprini},\ and\ \citenamefont
  {Lellouch}}]{Bourrely:2008za}%
  \BibitemOpen
  \bibfield  {author} {\bibinfo {author} {\bibfnamefont {C.}~\bibnamefont
  {Bourrely}}, \bibinfo {author} {\bibfnamefont {I.}~\bibnamefont {Caprini}}, \
  and\ \bibinfo {author} {\bibfnamefont {L.}~\bibnamefont {Lellouch}},\ }\href
  {\doibase 10.1103/PhysRevD.82.099902, 10.1103/PhysRevD.79.013008} {\bibfield
  {journal} {\bibinfo  {journal} {Phys. Rev.}\ }\textbf {\bibinfo {volume} {D
  79}},\ \bibinfo {pages} {013008} (\bibinfo {year} {2009})},\ \bibinfo {note}
  {[Erratum: Phys. Rev. D82, 099902 (2010)]},\ \Eprint
  {http://arxiv.org/abs/0807.2722} {arXiv:0807.2722 [hep-ph]} \BibitemShut
  {NoStop}%
\bibitem [{\citenamefont {Aoki}\ \emph {et~al.}(2020)\citenamefont {Aoki} \emph
  {et~al.}}]{Aoki:2019cca}%
  \BibitemOpen
  \bibfield  {author} {\bibinfo {author} {\bibfnamefont {S.}~\bibnamefont
  {Aoki}} \emph {et~al.} (\bibinfo {collaboration} {Flavour Lattice Averaging
  Group}),\ }\href {\doibase 10.1140/epjc/s10052-019-7354-7} {\bibfield
  {journal} {\bibinfo  {journal} {Eur. Phys. J. C}\ }\textbf {\bibinfo {volume}
  {80}},\ \bibinfo {pages} {113} (\bibinfo {year} {2020})},\ \Eprint
  {http://arxiv.org/abs/1902.08191} {arXiv:1902.08191 [hep-lat]} \BibitemShut
  {NoStop}%
\bibitem [{\citenamefont {Balitsky}\ \emph {et~al.}(1989)\citenamefont
  {Balitsky}, \citenamefont {Braun},\ and\ \citenamefont
  {Kolesnichenko}}]{BALITSKY1989509}%
  \BibitemOpen
  \bibfield  {author} {\bibinfo {author} {\bibfnamefont {I.}~\bibnamefont
  {Balitsky}}, \bibinfo {author} {\bibfnamefont {V.}~\bibnamefont {Braun}}, \
  and\ \bibinfo {author} {\bibfnamefont {A.}~\bibnamefont {Kolesnichenko}},\
  }\href {\doibase https://doi.org/10.1016/0550-3213(89)90570-1} {\bibfield
  {journal} {\bibinfo  {journal} {Nuclear Physics B}\ }\textbf {\bibinfo
  {volume} {312}},\ \bibinfo {pages} {509} (\bibinfo {year}
  {1989})}\BibitemShut {NoStop}%
\bibitem [{\citenamefont {Braun}\ and\ \citenamefont
  {Filyanov}(1989)}]{Braun:1989qcd}%
  \BibitemOpen
  \bibfield  {author} {\bibinfo {author} {\bibfnamefont {V.~M.}\ \bibnamefont
  {Braun}}\ and\ \bibinfo {author} {\bibfnamefont {I.~E.}\ \bibnamefont
  {Filyanov}},\ }\href {\doibase 10.1007/BF01548594} {\bibfield  {journal}
  {\bibinfo  {journal} {Zeitschrift f{\"u}r Physik C Particles and Fields}\
  }\textbf {\bibinfo {volume} {44}},\ \bibinfo {pages} {157} (\bibinfo {year}
  {1989})}\BibitemShut {NoStop}%
\bibitem [{\citenamefont {Chernyak}\ and\ \citenamefont
  {Zhitnitsky}(1990)}]{CHERNYAK1990137}%
  \BibitemOpen
  \bibfield  {author} {\bibinfo {author} {\bibfnamefont {V.}~\bibnamefont
  {Chernyak}}\ and\ \bibinfo {author} {\bibfnamefont {I.}~\bibnamefont
  {Zhitnitsky}},\ }\href {\doibase
  https://doi.org/10.1016/0550-3213(90)90612-H} {\bibfield  {journal} {\bibinfo
   {journal} {Nuclear Physics B}\ }\textbf {\bibinfo {volume} {345}},\ \bibinfo
  {pages} {137} (\bibinfo {year} {1990})}\BibitemShut {NoStop}%
\bibitem [{\citenamefont {Ball}\ \emph {et~al.}(1991)\citenamefont {Ball},
  \citenamefont {Braun},\ and\ \citenamefont {Dosch}}]{PhysRevD.44.3567}%
  \BibitemOpen
  \bibfield  {author} {\bibinfo {author} {\bibfnamefont {P.}~\bibnamefont
  {Ball}}, \bibinfo {author} {\bibfnamefont {V.~M.}\ \bibnamefont {Braun}}, \
  and\ \bibinfo {author} {\bibfnamefont {H.~G.}\ \bibnamefont {Dosch}},\ }\href
  {\doibase 10.1103/PhysRevD.44.3567} {\bibfield  {journal} {\bibinfo
  {journal} {Phys. Rev. D}\ }\textbf {\bibinfo {volume} {44}},\ \bibinfo
  {pages} {3567} (\bibinfo {year} {1991})}\BibitemShut {NoStop}%
\bibitem [{\citenamefont {Ball}\ and\ \citenamefont
  {Zwicky}(2005)}]{Ball:2004rg}%
  \BibitemOpen
  \bibfield  {author} {\bibinfo {author} {\bibfnamefont {P.}~\bibnamefont
  {Ball}}\ and\ \bibinfo {author} {\bibfnamefont {R.}~\bibnamefont {Zwicky}},\
  }\href {\doibase 10.1103/PhysRevD.71.014029} {\bibfield  {journal} {\bibinfo
  {journal} {Phys. Rev. D}\ }\textbf {\bibinfo {volume} {71}},\ \bibinfo
  {pages} {014029} (\bibinfo {year} {2005})},\ \Eprint
  {http://arxiv.org/abs/hep-ph/0412079} {arXiv:hep-ph/0412079} \BibitemShut
  {NoStop}%
\bibitem [{\citenamefont {Khodjamirian}\ \emph {et~al.}(2007)\citenamefont
  {Khodjamirian}, \citenamefont {Mannel},\ and\ \citenamefont
  {Offen}}]{Khodjamirian:2006st}%
  \BibitemOpen
  \bibfield  {author} {\bibinfo {author} {\bibfnamefont {A.}~\bibnamefont
  {Khodjamirian}}, \bibinfo {author} {\bibfnamefont {T.}~\bibnamefont
  {Mannel}}, \ and\ \bibinfo {author} {\bibfnamefont {N.}~\bibnamefont
  {Offen}},\ }\href {\doibase 10.1103/PhysRevD.75.054013} {\bibfield  {journal}
  {\bibinfo  {journal} {Phys. Rev. D}\ }\textbf {\bibinfo {volume} {75}},\
  \bibinfo {pages} {054013} (\bibinfo {year} {2007})},\ \Eprint
  {http://arxiv.org/abs/hep-ph/0611193} {arXiv:hep-ph/0611193} \BibitemShut
  {NoStop}%
\bibitem [{\citenamefont {Bharucha}(2012)}]{Bharucha:2012wy}%
  \BibitemOpen
  \bibfield  {author} {\bibinfo {author} {\bibfnamefont {A.}~\bibnamefont
  {Bharucha}},\ }\href {\doibase 10.1007/JHEP05(2012)092} {\bibfield  {journal}
  {\bibinfo  {journal} {JHEP}\ }\textbf {\bibinfo {volume} {05}},\ \bibinfo
  {pages} {092} (\bibinfo {year} {2012})},\ \Eprint
  {http://arxiv.org/abs/1203.1359} {arXiv:1203.1359 [hep-ph]} \BibitemShut
  {NoStop}%
\bibitem [{\citenamefont {Freytsis}\ \emph {et~al.}(2015)\citenamefont
  {Freytsis}, \citenamefont {Ligeti},\ and\ \citenamefont
  {Ruderman}}]{Freytsis:2015qca}%
  \BibitemOpen
  \bibfield  {author} {\bibinfo {author} {\bibfnamefont {M.}~\bibnamefont
  {Freytsis}}, \bibinfo {author} {\bibfnamefont {Z.}~\bibnamefont {Ligeti}}, \
  and\ \bibinfo {author} {\bibfnamefont {J.~T.}\ \bibnamefont {Ruderman}},\
  }\href {\doibase 10.1103/PhysRevD.92.054018} {\bibfield  {journal} {\bibinfo
  {journal} {Phys. Rev. D}\ }\textbf {\bibinfo {volume} {92}},\ \bibinfo
  {pages} {054018} (\bibinfo {year} {2015})},\ \Eprint
  {http://arxiv.org/abs/1506.08896} {arXiv:1506.08896 [hep-ph]} \BibitemShut
  {NoStop}%
\bibitem [{\citenamefont {Bernlochner}\ and\ \citenamefont
  {Ligeti}(2017)}]{Bernlochner:2016bci}%
  \BibitemOpen
  \bibfield  {author} {\bibinfo {author} {\bibfnamefont {F.~U.}\ \bibnamefont
  {Bernlochner}}\ and\ \bibinfo {author} {\bibfnamefont {Z.}~\bibnamefont
  {Ligeti}},\ }\href {\doibase 10.1103/PhysRevD.95.014022} {\bibfield
  {journal} {\bibinfo  {journal} {Phys. Rev. D}\ }\textbf {\bibinfo {volume}
  {95}},\ \bibinfo {pages} {014022} (\bibinfo {year} {2017})},\ \Eprint
  {http://arxiv.org/abs/1606.09300} {arXiv:1606.09300 [hep-ph]} \BibitemShut
  {NoStop}%
\bibitem [{\citenamefont {Dor\v{s}ner}\ \emph {et~al.}(2016)\citenamefont
  {Dor\v{s}ner}, \citenamefont {Fajfer}, \citenamefont {Greljo}, \citenamefont
  {Kamenik},\ and\ \citenamefont {Ko\v{s}nik}}]{Dorsner:2016wpm}%
  \BibitemOpen
  \bibfield  {author} {\bibinfo {author} {\bibfnamefont {I.}~\bibnamefont
  {Dor\v{s}ner}}, \bibinfo {author} {\bibfnamefont {S.}~\bibnamefont {Fajfer}},
  \bibinfo {author} {\bibfnamefont {A.}~\bibnamefont {Greljo}}, \bibinfo
  {author} {\bibfnamefont {J.~F.}\ \bibnamefont {Kamenik}}, \ and\ \bibinfo
  {author} {\bibfnamefont {N.}~\bibnamefont {Ko\v{s}nik}},\ }\href {\doibase
  10.1016/j.physrep.2016.06.001} {\bibfield  {journal} {\bibinfo  {journal}
  {Phys. Rept.}\ }\textbf {\bibinfo {volume} {641}},\ \bibinfo {pages} {1}
  (\bibinfo {year} {2016})},\ \Eprint {http://arxiv.org/abs/1603.04993}
  {arXiv:1603.04993 [hep-ph]} \BibitemShut {NoStop}%
\bibitem [{\citenamefont {Buchmüller}\ \emph {et~al.}(1987)\citenamefont
  {Buchmüller}, \citenamefont {Rückl},\ and\ \citenamefont
  {Wyler}}]{BUCHMULLER1987442}%
  \BibitemOpen
  \bibfield  {author} {\bibinfo {author} {\bibfnamefont {W.}~\bibnamefont
  {Buchmüller}}, \bibinfo {author} {\bibfnamefont {R.}~\bibnamefont {Rückl}},
  \ and\ \bibinfo {author} {\bibfnamefont {D.}~\bibnamefont {Wyler}},\ }\href
  {\doibase https://doi.org/10.1016/0370-2693(87)90637-X} {\bibfield  {journal}
  {\bibinfo  {journal} {Physics Letters B}\ }\textbf {\bibinfo {volume}
  {191}},\ \bibinfo {pages} {442} (\bibinfo {year} {1987})}\BibitemShut
  {NoStop}%
\bibitem [{\citenamefont {Lange}(2001)}]{Lange:2001uf}%
  \BibitemOpen
  \bibfield  {author} {\bibinfo {author} {\bibfnamefont {D.~J.}\ \bibnamefont
  {Lange}},\ }\bibfield  {booktitle} {\emph {\bibinfo {booktitle}
  {{Proceedings, 7th International Conference on B physics at hadron machines
  (BEAUTY 2000): Maagan, Israel, September 13-18, 2000}}},\ }\href {\doibase
  10.1016/S0168-9002(01)00089-4} {\bibfield  {journal} {\bibinfo  {journal}
  {Nucl. Instrum. Meth.}\ }\textbf {\bibinfo {volume} {A462}},\ \bibinfo
  {pages} {152} (\bibinfo {year} {2001})}\BibitemShut {NoStop}%
\bibitem [{\citenamefont {Gubernari}\ \emph {et~al.}(2019)\citenamefont
  {Gubernari}, \citenamefont {Kokulu},\ and\ \citenamefont {van
  Dyk}}]{Gubernari:2018wyi}%
  \BibitemOpen
  \bibfield  {author} {\bibinfo {author} {\bibfnamefont {N.}~\bibnamefont
  {Gubernari}}, \bibinfo {author} {\bibfnamefont {A.}~\bibnamefont {Kokulu}}, \
  and\ \bibinfo {author} {\bibfnamefont {D.}~\bibnamefont {van Dyk}},\ }\href
  {\doibase 10.1007/JHEP01(2019)150} {\bibfield  {journal} {\bibinfo  {journal}
  {JHEP}\ }\textbf {\bibinfo {volume} {01}},\ \bibinfo {pages} {150} (\bibinfo
  {year} {2019})},\ \Eprint {http://arxiv.org/abs/1811.00983} {arXiv:1811.00983
  [hep-ph]} \BibitemShut {NoStop}%
\bibitem [{\citenamefont {Tanaka}\ and\ \citenamefont
  {Watanabe}(2013)}]{Tanaka:2012nw}%
  \BibitemOpen
  \bibfield  {author} {\bibinfo {author} {\bibfnamefont {M.}~\bibnamefont
  {Tanaka}}\ and\ \bibinfo {author} {\bibfnamefont {R.}~\bibnamefont
  {Watanabe}},\ }\href {\doibase 10.1103/PhysRevD.87.034028} {\bibfield
  {journal} {\bibinfo  {journal} {Phys. Rev.}\ }\textbf {\bibinfo {volume}
  {D87}},\ \bibinfo {pages} {034028} (\bibinfo {year} {2013})},\ \Eprint
  {http://arxiv.org/abs/1212.1878} {arXiv:1212.1878 [hep-ph]} \BibitemShut
  {NoStop}%
\bibitem [{\citenamefont {Ligeti}\ \emph {et~al.}(2017)\citenamefont {Ligeti},
  \citenamefont {Papucci},\ and\ \citenamefont {Robinson}}]{Ligeti:2016npd}%
  \BibitemOpen
  \bibfield  {author} {\bibinfo {author} {\bibfnamefont {Z.}~\bibnamefont
  {Ligeti}}, \bibinfo {author} {\bibfnamefont {M.}~\bibnamefont {Papucci}}, \
  and\ \bibinfo {author} {\bibfnamefont {D.~J.}\ \bibnamefont {Robinson}},\
  }\href {\doibase 10.1007/JHEP01(2017)083} {\bibfield  {journal} {\bibinfo
  {journal} {JHEP}\ }\textbf {\bibinfo {volume} {01}},\ \bibinfo {pages} {083}
  (\bibinfo {year} {2017})},\ \Eprint {http://arxiv.org/abs/1610.02045}
  {arXiv:1610.02045 [hep-ph]} \BibitemShut {NoStop}%
\bibitem [{\citenamefont {Bernlochner}\ \emph
  {et~al.}(2018{\natexlab{a}})\citenamefont {Bernlochner}, \citenamefont
  {Ligeti},\ and\ \citenamefont {Robinson}}]{Bernlochner:2017jxt}%
  \BibitemOpen
  \bibfield  {author} {\bibinfo {author} {\bibfnamefont {F.~U.}\ \bibnamefont
  {Bernlochner}}, \bibinfo {author} {\bibfnamefont {Z.}~\bibnamefont {Ligeti}},
  \ and\ \bibinfo {author} {\bibfnamefont {D.~J.}\ \bibnamefont {Robinson}},\
  }\href {\doibase 10.1103/PhysRevD.97.075011} {\bibfield  {journal} {\bibinfo
  {journal} {Phys. Rev.}\ }\textbf {\bibinfo {volume} {D97}},\ \bibinfo {pages}
  {075011} (\bibinfo {year} {2018}{\natexlab{a}})},\ \Eprint
  {http://arxiv.org/abs/1711.03110} {arXiv:1711.03110 [hep-ph]} \BibitemShut
  {NoStop}%
\bibitem [{\citenamefont {Bernlochner}\ \emph
  {et~al.}(2018{\natexlab{b}})\citenamefont {Bernlochner}, \citenamefont
  {Ligeti}, \citenamefont {Robinson},\ and\ \citenamefont
  {Sutcliffe}}]{Bernlochner:2018kxh}%
  \BibitemOpen
  \bibfield  {author} {\bibinfo {author} {\bibfnamefont {F.~U.}\ \bibnamefont
  {Bernlochner}}, \bibinfo {author} {\bibfnamefont {Z.}~\bibnamefont {Ligeti}},
  \bibinfo {author} {\bibfnamefont {D.~J.}\ \bibnamefont {Robinson}}, \ and\
  \bibinfo {author} {\bibfnamefont {W.~L.}\ \bibnamefont {Sutcliffe}},\ }\href
  {\doibase 10.1103/PhysRevLett.121.202001} {\bibfield  {journal} {\bibinfo
  {journal} {Phys. Rev. Lett.}\ }\textbf {\bibinfo {volume} {121}},\ \bibinfo
  {pages} {202001} (\bibinfo {year} {2018}{\natexlab{b}})},\ \Eprint
  {http://arxiv.org/abs/1808.09464} {arXiv:1808.09464 [hep-ph]} \BibitemShut
  {NoStop}%
\end{thebibliography}%

\onecolumngrid

\end{document}